\documentclass[pra,twocolumn,preprintnumbers,amsmath,amssymb,superscriptaddress,showpacs,longbibliography]{revtex4-1}
\usepackage{graphicx}
\usepackage{amsmath}
\usepackage{graphics}
\usepackage{amssymb}
\usepackage{amsfonts,epsfig}
\usepackage{natbib}
\usepackage{hyperref}
\usepackage{enumitem}
\usepackage{dsfont}

\allowdisplaybreaks

\newcommand{ \id }
							{d}
\newcommand{ \Fc }[1]{ c_{#1} }								

\newcommand{ \dchi }{ \Delta\chi }								
\newcommand{ \chispec}{ \chi_S }								
\newcommand{ \exec }{execution}								
\newcommand{ \method }{methodology}						
\newcommand{ \sinc }{ \mathrm{sinc} }						
\newcommand{ \erf }{ \mathrm{erf} }							
\newcommand{ \g }{ K }											
\newcommand{ \f }{ J }											
\newcommand{ \as }{\'{A}lvarez-Suter}
\newcommand{ \ASM }
							{\'{A}lvarez-Suter method}

\usepackage{color}

\usepackage{ulem} 

\begin{document}

\title{Accuracy of dynamical-decoupling-based spectroscopy of Gaussian noise}
\author{Piotr Sza\'{n}kowski}\email{piotr.szankowski@ifpan.edu.pl}
\affiliation{Institute of Physics, Polish Academy of Sciences, al.~Lotnik{\'o}w 32/46, PL 02-668 Warsaw, Poland}
\author{{\L}ukasz Cywi{\'n}ski}
\affiliation{Institute of Physics, Polish Academy of Sciences, al.~Lotnik{\'o}w 32/46, PL 02-668 Warsaw, Poland}

\date{\today}

\begin{abstract}
The fundamental assumption of dynamical decoupling based noise spectroscopy is that the coherence decay rate of qubit (or qubits) driven with a sequence of many pulses, is well approximated by the environmental noise spectrum spanned on frequency comb defined by the sequence. Here we investigate the precise conditions under which this commonly used spectroscopic approach is quantitatively correct. To this end we focus on two representative examples of spectral densities: the long-tailed Lorentzian, and finite-ranged Gaussian---both expected to be encountered when using the qubit for nano-scale nuclear resonance imaging. We have found that, in contrast to Lorentz spectrum, for which the corrections to the standard spectroscopic formulas can easily be made negligible, the spectra with finite range are more challenging to reconstruct accurately. For Gaussian line-shape of environmental spectral density, direct application of the standard dynamical decoupling based spectroscopy leads to erroneous attribution of long-tail behavior to the reconstructed spectrum. Fortunately, artifacts such as this, can be completely avoided with the simple extension to standard reconstruction method.
\end{abstract}

\maketitle

\section{Introduction}\label{sec:Introduction}
 
 The dynamical decoupling based noise spectroscopy (DDNS) is a powerful tool for sensing and analyzing the signal emitted by the environment of a coherently controlled probe composed of single \cite{Szankowski_JPCM17,Degen_RMP17,Bylander_NP11,Alvarez_PRL11,Yuge_PRL11,Kotler_Nature11}, or multiple \cite{Szankowski_PRA16,Paz_NJP16,Paz_PRA17} qubits. It has been implemented with essentially all kinds of qubits \cite{Biercuk_Nature09,Bylander_NP11,Alvarez_PRL11,Kotler_Nature11,Almog_JPB11,Staudacher_Science13,Medford_PRL12,Muhonen_NN14,BarGill_NC12,Romach_PRL15,Malinowski_PRL17}, with arguably the most promising direction of recent research focused on application of spin qubits in semiconductors (e.g. nitrogen-vacancy centers in diamond \cite{Rondin_RPP14,Wrachtrup_JMR16}) as a nanoscale magnetometers of nuclear spin environmental noise \cite{Staudacher_Science13,Lovchinsky_Science16}. 

The single qubit DDNS method utilizes a two level probe coupled to its environment via a phase noise Hamiltonian $\hat H_\mathrm{int}\propto\hat{V}\hat\sigma_z$. Here the Pauli operator $\hat{\sigma}_{z}$ pertains to the qubit, while the environmental operator $\hat{V}$ can be replaced with a stochastic function when the semi-classical approximation is applicable to environment-qubit coupling. It is assumed that the coupling does not facilitate energy intake from the environmental noise (i.e. the coupling operator commutes with the free Hamiltonian of the qubit, $[\hat H_Q , \hat H_\mathrm{int}]=0$), so that the interaction only affects the coherence between $\hat\sigma_z$ operator eigenstates $|\pm\rangle$. Ultimately, it leads to the decay of this coherence---this is the process of pure dephasing, a specific kind of decoherence. The course of this decoherence process is directly influenced by the properties of the noise.  Hence, it is possible to infer those properties from the examination of the qubit's evolution. However, a passive observation offers only limited insight, and in order to obtain more detailed quantitative information one needs a method for signal analysis. This is accomplished by controlling the qubit with a sequences of virtually instantaneous $\pi$ pulses that cause its Bloch vector to flip, without introducing any other kind of disruption to the normal phase evolution. 
The properly devised pulse sequences act as an adjustable narrow-band frequency filter of the incoming noise \cite{Szankowski_JPCM17,Biercuk_JPB11,Martinis_PRB03, Bylander_NP11,Ajoy_PRA11,Alvarez_PRL11,Yuge_PRL11,Kotler_Nature11,Almog_JPB11}.
Originally, this method was used to decouple the qubit from the environment (and in this way enhance its coherence time) by setting the passband of the filter beyond the frequency range of the noise \cite{Viola_PRA98,Viola_PRL99,Uhrig_PRL07,Gordon_PRL08,Cywinski_PRB08,Khodjasteh_NC13,Suter_RMP16}. Because of this initial application, the qubit control via pulse sequences is often referred to as dynamical decoupling. In the context of noise spectroscopy, it allows one to dissect the spectral distribution of the signal by decoupling the qubit from all but a handful of frequencies, thus relating the dephasing rate only to those specific parts of the noise spectrum \cite{Alvarez_PRL11,Bylander_NP11,Yuge_PRL11,Biercuk_JPB11,Almog_JPB11,Szankowski_JPCM17}.

The standard theoretical formulation of the noise spectroscopy method relies on the approximation where the width of the passband of the pulse sequence-induced frequency filter is set to zero. In practice, this approximation amounts to replacing the band-pass filter function that modulates the noise, with a series of Dirac deltas (so-called Dirac comb) centered at frequencies defined by the properties of the applied pulse sequence \cite{deSousa_TAP09,Cywinski_PRB08,Biercuk_JPB11,Szankowski_JPCM17,Bylander_NP11,Alvarez_PRL11,Yuge_PRL11}. The resultant expression for the qubit's decoherence rate, to which we will refer to as \textit{the spectroscopic formula}, establishes an invertible relation with the values of noise spectrum, enabling their reconstruction from the measured coherence lifetimes. The main goal of this paper is to quantitatively understand the behavior of the corrections to this approximate result, and consequently to elucidate the conditions under which they become negligible---in other words, to quantitatively define the \textit{spectroscopic regime} of evolution parameters. To this end we will examine the {\it exact} results for the decoherence and compare it with the spectroscopic formulas. It goes without saying that these kinds of calculations are not feasible for completely arbitrary noise spectrum. Instead, we consider two representative types of noise spectra, relevant for nanoscale nuclear sensing \cite{Staudacher_Science13,Muller_NC14,DeVience_NN15,Haberle_NN15,Lovchinsky_Science16,Boss_Science17,Schmitt_Science17}: (i) a spectral density with a {\it long tail}, i.e. having a power-law decay for large frequencies ($\sim (1/\omega)^\beta$ as $\omega\to\infty$, with $\beta\geqslant 2$ and $\beta\in\mathrm{Integers}$), exemplified by the Lorentzian spectrum, (ii) the {\it finite-range} Gaussian-shaped spectrum, that decays faster than power-law \cite{Abragam}. The most important, from practical point of view, conclusion of our analysis is that the noise spectroscopy method based on the assumption of operating within a delta-function approximation leads to artifacts in reconstruction of tails of Gaussian-shaped spectrum (more generally, any spectrum with finite range). The basic reason for this is the following. The peaks of the frequency filter generated by the pulse sequence are of course only approximately delta-shaped---in fact the envelope of the filter decays as a power-law as a function of detuning from one of its maxima. The overlap between these tails of the filter envelope and the spectrum is the main source of corrections to the spectroscopic formula. The crucial, and somewhat obvious, observation regarding the finite-ranged type spectrum is that when the characteristic frequency of the sequence is set outside of its range, the spectral density  at this frequency  becomes negligibly small, which in turn makes the ``spectroscopic'' contribution negligible as well. When this is the case, the {\it only} contribution to dephasing is due to the above-mentioned corrections. In those circumstances, when one operates under an assumption that the delta-function approximation holds true, one is forced to mistakingly interpret the correction as a feature of the reconstructed spectrum. While a similar problem might also arise in the case of long-tailed spectrum, the key difference is that the dependencies of the spectroscopic formula and the correction on the ``filtering frequency'' of the sequence are now of the same character---both exhibit power-law tails, which makes the relative error much easier to manage with adjustments to used pulse sequences. Unfortunately, in the case of finite-ranged spectra, no feasible modifications of pulse sequences could overcome this problem. However, we show that a simple extension to the reconstruction method allows one to obtain results completely free of these artifacts, even if the relative error between spectroscopic formula and the correction is large.

The structure of the paper is the following. In Sec. \ref{sec:statement} we briefly summarize the theoretical underpinnings of DDNS, and the possible sources of inaccuracies of spectrum reconstruction. Then, in Sec. \ref{sec:spec_formula}, we elucidate the structure of the exact result for decoherence due to Gaussian phase noise, and identify the expressions corresponding to the spectroscopic regime, and the correction terms. The calculations for Lorentzian- and Gaussian-shaped spectra are presented in Sec.\ref{sec:lorentzian} and \ref{sec:gauss}, respectively, the latter section ending with discussion of erroneous attribution of the long tail to the finite-ranged spectrum when the standard reconstruction method is applied. We show how to fix this problem with an appropriate modification to the standard method introduced in Sec.~\ref{sec:fix}. Finally, in Sec.~\ref{sec:cut-off_error} we discuss the other sources of reconstruction errors (due to use of a finite number of measurements) for the modified spectroscopic method.

\section{Statement of the theoretical problem} \label{sec:statement}
In technical terms, the phase noise driving the evolution of the qubit imprints some information about its properties onto the off-diagonal elements of the qubit density matrix---its coherence
\begin{equation}
W(T) = \frac{\langle {+}|\hat\rho_Q(T)|{-}\rangle}{\langle {+}|\hat\rho_Q(0)|{-}\rangle} \equiv e^{-\chi(T)}\,. \label{eq:chi}
\end{equation}
Here, $|\pm\rangle$ are the eigenstates of the $Z$-component of the qubit spin operator, and $\hat\rho_Q$ is the density matrix of the qubit obtained by averaging over noise realizations in the case of classical noise, or by partial tracing the full qubit-environment state. The exponent above is called the attenuation function, and for stationary Gaussian noise is determined by the noise correlation function $C(|t|)$ (this is true for both classical and quantum noise \cite{Szankowski_JPCM17,Paz_PRA17}),
\begin{equation}
\chi(T) = \frac{1}{2}\int_0^T \id t \int_0^T \id t' C(|t-t'|)f_T(t)f_T(t')\,,\label{eq:attfc}
\end{equation}
where the time-domain filter function,
\begin{equation}
f_T(t) = \sum_{k=0}^n (-1)^k \Theta\left(t_{k+1}-t\right)\Theta\left(t-t_k\right)\,,
\end{equation}
encapsulates the effects of the control sequence composed of $n$ $\pi$-pulses applied to the qubit at times $t_1,t_2,\ldots,t_n$  (we also used the conventional notation where $t_0=0$ and $t_{n+1}=T$).

In the spectroscopic regime (discussed in greater detail in Sec. \ref{sec:spec_formula}) the filter functions are approximated by the frequency Dirac comb and the imprinted information consists of the noise power spectrum, $S(\omega)=\int_{-\infty}^\infty e^{-i\omega t}C(|t|)\id t$, encoded in the decoherence rate via {\it spectroscopic formula},
\begin{equation}
-\ln W(T) = \chi(T) \approx T\sum_{m>0}|\Fc{m\omega_p}|^2 S(m\omega_p)\,. \label{eq:spec_formula}
\end{equation}
(Note the absence of zero frequency term in the sum.) The coefficients weighting the spectrum are the Fourier series coefficients of the filter,
\begin{align}
f_T(t) &=\Theta\left(T-t\right)\Theta\left(t\right) \sum_{m\neq 0} \Fc{m\omega_p} e^{i m\omega_p t}\,,\label{eq:F_series}\\
\Fc{\omega} &= \frac{1}{T}\int_0^T e^{-i \omega t}f_T(t) \id t\,,\label{eq:f_T}
\end{align}
and the characteristic frequency of the sequence, $\omega_p$, is the smallest multiple of $2\pi/T$ present in the expansion \cite{Szankowski_JPCM17}. The lack of zero frequency term in Eq.~\eqref{eq:spec_formula} is a consequence of an additional assumption that all pulse sequences considered here are ``balanced'' , i.e. $\Fc{0}=T^{-1}\int_0^T f_T(t)\id t = 0$.

Assuming the spectroscopic formula is exact, the function $S(\omega)$ can now be reconstructed using the \as{} method \cite{Alvarez_PRL11,Szankowski_JPCM17}. The necessary data is acquired in a series of experiments, each performed with a different pulse sequence characterized by frequency $\omega_j$ and Fourier coefficients $\Fc{m\omega_j}^{(j)}$. Then, the relationship between the corresponding measured attenuation functions and the unknown values of spectral density can be cast into vector form,
\begin{equation}\label{eq:AS_eq_sys}
\frac{\chi^{(j)}(T_j)}{T_j} \approx \sum_{m\leqslant m_c}|\Fc{m\omega_j}^{(j)}|^2 S(m\omega_j) \equiv \sum_{m} U_{j,m}S_m\,.
\end{equation}
Here, $U_{j,m}$ is a matrix composed of {\it known} Fourier coefficients, $S_m$ are the components of the vector of spectrum values picked out by the frequency combs of each applied sequence. Because of the frequency cut-off, $m_c$, introduced so that $\mathbf U$ is of finite dimension, it becomes possible to invert the above relation and restore the noise spectrum, $S_m = \sum_j U^{-1}_{m,j}\chi^{(j)}(T_j)/T_j$. Of course, the inverse of $U_{j,m}$ exists only if sequences used in the procedure are chosen in an appropriate way, e.g., the number of sequences (indexed with $j$) must be compatible with cut-off $m_c$, so that $\mathbf U$ is a square matrix.

Clearly, the accuracy of the reconstruction scheme is directly linked to the veracity of the spectroscopic formula itself. The disparity between the real value of the measured attenuation function and Eq.~\eqref{eq:spec_formula} has two main sources: the \exec{} and the \method{} errors.

The \exec{} errors include all ``technical'' shortcomings of the procedure: (i) the measurement error, and (ii) the discrepancy between the actual physical realization of $\pi$-pulse and their idealization that leads to filter function \eqref{eq:f_T}. From this point we will assume that the procedure is being executed perfectly and the above errors are negligibly small. Such an assumption is well motivated, since in many noise spectroscopy experiments \cite{deLange_Science10,Staudacher_Science13} the systematic pulse errors are being efficiently neutralized by routinely employed countermeasures \cite{Gullion_JMR90}. However, recently it was noticed that non negligible pulse durations might lead to an appearance of a modified filter function, for which the contributions from higher harmonics of $\omega_p$ are amplified \cite{Loretz_PRX15}. In certain circumstances, this can complicate the reconstruction of spectra of a particular structure \cite{Loretz_PRX15}. This can be counteracted by replacing short pulses with appropriately optimized continuous driving \cite{Frey_arXiv17}. Here we neglect this kind of error, as we focus on single-peaked spectral densities the results for which should not be strongly affected by finite pulse width.
 
The \method{} errors account for inaccurate statements made during the development of the theoretical side of the problem, namely:
\begin{enumerate}
\item The presence of transverse coupling that deviates the qubit evolution from the pure dephasing model.
\item Possible non-Gaussian nature of the noise, which causes the ``pollution'' of attenuation function \eqref{eq:attfc} with contribution from higher order noise correlations (noise cumulants).
\item In practice, the theoretical value of the coherence  given by Eq.~\eqref{eq:chi} cannot be acquired directly, but rather it must be estimated from a finite number of measurements. This limitation unavoidably leads to an imperfect estimation of the attenuation function~\eqref{eq:attfc}.
\item The deviation from frequency comb approximation to filter functions.
\item The introduction of cut-off $m_c$ required by the spectrum reconstruction method discussed above. 
\end{enumerate}
Here we shall assume that the noise is exactly Gaussian and the contribution from transverse couplings is negligible, so that the first and second point of the list can be completely ignored. As a side note, DDNS methods can be adapted to characterize the spectrum of transversely coupled low-frequency noise \cite{Cywinski_PRA14}, and a proposition for a general extension to DDNS enabling an analog of spectroscopy for the case of non-Gaussian noises has been recently reported in \cite{Norris_PRL16}. Moreover, we assume that the inaccuracies in the estimation of the attenuation function described in the third point, can be also safely neglected. The Bayesian approach to analysis of such finite sets of data was eloquently advocated for in the context of noise spectroscopy in \cite{Ferrie_arXiv17}.

 The fourth point is the main focus of the investigations presented in the following sections. The fifth item of the list will be addressed in Sec.~\ref{sec:cut-off_error}, where we will discuss the cut-off error of the modified \ASM{} (introduced in Sec.~\ref{sec:fix}).

\section{The spectroscopic regime}\label{sec:spec_formula}

The foundation of DDNS method, the spectroscopic formula \eqref{eq:spec_formula}, is an approximate form of the attenuation function \eqref{eq:attfc}. When the pulse sequence parameters (i.e., the duration $T$, and the characteristic frequency $\omega_p$) allow for approximating the attenuation function by Eq.~(\ref{eq:spec_formula}) with desired accuracy, we then operate in the spectroscopic regime. The main purpose of this section is to define this regime. Anticipating the structure of the final result we start by substituting in \eqref{eq:attfc} the time-domain filter functions with their Fourier series expansions \eqref{eq:F_series},
\begin{widetext}\begin{align}\label{eq:attfc_time}
\chi(T) &= \frac 12 \int_0^T \id t_1 \id t_2\, f_T(t_1)f_T(t_2)C(|t_1-t_2|)= \int_0^T \id t_1 \int_0^{t_1}\id t_2\, f_T(t_1)f_T(t_2)C(|t_1-t_2|)\nonumber\\
&= \sum_{m_1,m_2}\Fc{m_1\omega_p}\Fc{m_2\omega_p} \int_0^T \id \tau\, C(|\tau|) e^{i \frac{m_1-m_2}{2}\omega_p \tau}
	\int_{\frac{\tau}{2}}^{T-\frac \tau 2} \id\bar{t}\, e^{i(m_1+m_2)\omega_p\bar{t}}\nonumber\\
&=T\sum_{m}|\Fc{m\omega_p}|^2 \int_0^T \id \tau\,\left(1-\frac{\tau}T\right) C(|\tau|)e^{i m\omega_p \tau} 
	+ \sum_{m_1\neq -m_2} \frac{\Fc{m_1\omega_p}\Fc{m_2\omega_p}}{(m_1+m_2)\omega_p}
		\int_0^T \id \tau\, C(|\tau|)(e^{i m_1\omega_p \tau}-e^{-i m_2 \omega_p \tau})\,.
\end{align}\end{widetext}
The attenuation function has been split into two parts: the ``diagonal'' part that is proportional to $T$, and the ``off-diagonal'' remainder. 

In a nutshell, the main objective of the spectroscopic approximation is to extend the upper limit of integration in \eqref{eq:attfc_time} to infinity. By doing so the time integrals become Fourier transforms of $C$, thus providing the access to parts of noise spectrum at frequencies commensurate with the characteristic frequency of the sequence $\omega_p$. This objective can be achieved when, for fixed $\omega_p$, the sequence duration is much longer than the noise {\it correlation time} $\tau_c$, which quantifies the range of the correlation function, i.e., such $\tau_c$ that $C(\tau)\xrightarrow{\tau\gg \tau_c}0$. This decay of correlation function is expected on physical grounds. For a completely deterministic signal the correlation time is infinite, because it is possible to determine with certainty the value of the signal at one time point knowing its value at a different time (via its equations of motion). With the introduction of randomness the correlation time is shortened, as the ability of the observer to predict the value of the signal from its earlier state or to deduce its previous values from a given state, is diminished.

When $T\gg \tau_c$ the upper limit may just as well be infinity, because the integration effectively terminates at $\tau_c$ as the integrand vanishes beyond that point. Moreover, in the diagonal term we have $1-\tau/T\approx 1$, since under the integral $\tau \lesssim \tau_c \ll T$. Also, in the same regime, the off-diagonal remainder term can be dropped because $|\int_0^\infty \id \tau \, C(|\tau|)e^{i m\omega_p \tau}|\leq \int_0^\infty \id \tau |C(|\tau|)| \propto \tau_c$, which is much smaller than the $T$-scaling diagonal term. In summary, we have
\begin{align}
\chi(T) &{}\xrightarrow{T\gg \tau_c}T\sum_{m} |\Fc{m\omega_p}|^2 \int_0^\infty \id \tau\, C(|\tau|)e^{-i m\omega_p \tau}\nonumber\\
&=T\sum_{m>0}\left[|\Fc{m\omega_p}|^2 \int_0^\infty \id \tau\, C(|\tau|)e^{-i m\omega_p \tau}\right.\nonumber\\
&\phantom{T\sum_{m>0}}\left.+|\Fc{-m\omega_p}|^2 \int_0^\infty \id \tau\, C(|\tau|)e^{+i m\omega_p \tau}\right]\nonumber\\
&=T\sum_{m>0}|\Fc{m\omega_p}|^2\Big(\! \int_0^\infty \!\!+\!\! \int_{-\infty}^0\!\Big)\id \tau\, C(|\tau|)e^{-im\omega_p \tau}\nonumber\\
&=T\sum_{m>0} |\Fc{m\omega_p}|^2 S(m\omega_p)\,,
\end{align}
so that the spectroscopic formula is obtained asymptotically for long sequence durations. 

An alternative and more popular approach is to first transform the attenuation function to frequency space,
\begin{equation}
\chi(T) = \frac{1}{2}\int_{-\infty}^\infty |\tilde{f}_T(\omega)|^2 S(\omega)\frac{\id \omega}{2\pi}\,,\label{eq:attfc_freq}
\end{equation}
where $\tilde{f}_T(\omega)$ is the Fourier transform of $f_T(t)$. For large $T$ the frequency-domain filter $|\tilde{f}_T(\omega)|^2$ has a form of sharply multi-peaked function, since
\begin{subequations}\label{eq:freq_dom_filter}\begin{align}
&|\tilde{f}_T(\omega)|^2 = \left|\int_{-\infty}^\infty \!\!\!\id t\,e^{-i \omega t}f_T(t)\right|^2\nonumber\\
& = \left|\int_0^T \!\!\id t\,e^{-i\omega t}\left(\sum_m \Fc{m\omega_p}e^{i m\omega_p t}\right)\right|^2\nonumber\\
\nonumber
&=\left|\sum_m \Fc{m\omega_p} e^{-i\frac{T(\omega-m\omega_p)}{2}}T\, \mathrm{sinc}\left( \frac{T(\omega-m\omega_p)}{2}\right)\right|^2\\[.3cm]
\label{subeq:ff_diag}
&	= T\sum_{m}|\Fc{m\omega_p}|^2\,T\,\mathrm{sinc}^2\left[\frac{T(\omega-m\omega_p)}{2}\right]\\
\nonumber
&+\sum_{m_1\neq m_2}\Fc{m_1\omega_p}\Fc{m_2\omega_p}^*e^{i \omega_p T \left(\frac{m_1-m_2}{2}\right)}\times\\
\label{subeq:ff_off-diag}
&	 T\,\mathrm{sinc}\left[\frac{T(\omega-m_1\omega_p)}{2}\right]T\,\mathrm{sinc}\left[\frac{T(\omega-m_2\omega_p)}{2}\right]\,.
\end{align}\end{subequations}
The Dirac comb structure is obtained by passing to the limit of (infinitely) long sequence duration, 
\begin{align}
&|\tilde{f}_T(\omega)|^2\xrightarrow{T\to\infty}  2\pi T\sum_{m}|\Fc{m\omega_p}|^2 \delta(\omega-m\omega_p)\nonumber\\
&	+4\pi^2\sum_{m_1\neq m_2}\Fc{m_1\omega_p} \Fc{m_2\omega_p}^*\delta(\omega-m_1\omega_p)\delta(\omega-m_2\omega_p)\nonumber\\
&= 2\pi T\sum_m |\Fc{m\omega_p}|^2 \delta(\omega-m\omega_p)\,.\label{eq:comb_approx}
\end{align}
Of course, strictly speaking, the limit $T\to \infty$ is unachievable, and instead $T$ should be compared with some characteristic time scale associated with $S(\omega)$. Usually it is argued that the spectral density changes significantly on frequency scales of order $\tau_c^{-1}$. Therefore, provided that $T\gg \tau_c$, the rate of change of the spectrum is much slower than $\mathrm{sinc}$ functions, so that $S(\omega)$ can be treated as a constant around each peak. In this regime, each peak can be simply substituted with a delta function like in Eq.~\eqref{eq:comb_approx}, even for finite $T$. Inserting this form of filter function into Eq.~\eqref{eq:attfc_freq} yields the spectroscopic formula.

As it was touched upon previously, the conditions for achieving a spectroscopic regime should be investigated in the settings where $\omega_p$ is fixed while the duration of the sequence $T$ is treated as an adjustable parameter. Furthermore, by adopting an additional constrain that the sequence duration is manipulated only by increasing or decreasing the number of repetitions of a given pulse sequence, also the Fourier coefficients $\Fc{m\omega_p}$ become independent of $T$ and can be considered as fixed. To verify this assertion, let us define $f_{T_B}(t)$ as the time-domain filter function of the ``base'' sequence to be repeated, with duration $T_B$ and characteristic frequency $\omega_p$. Keeping in mind that $\omega_p$ is by definition a multiple of $2\pi/T_B$, we get
\begin{align}
\nonumber
&\Fc{m\omega_p} = \frac{1}{T_B}\int_0^{T_B} \id t\, f_{T_B}(t)e^{-i m \omega_p t} \frac{n}{n}\\
\nonumber
	 &= \frac{1}{T_B} \int_{0}^{T_B} \id t f_{T_B}(t)e^{-i m \omega_p t}\frac{1}{n}\sum_{k=0}^{n-1}e^{i T_B k m \omega_p}\\
\nonumber
	 &= \frac{1}{n T_B} \sum_{k=0}^{n-1} \int_{k T_B}^{(k+1)T_B} \!\!\!\id t\, f_{T_B}(t-k T_B)e^{-i m\omega_p t}\\
	 &=\frac{1}{n T_B} \int_0^{n T_B} \id t\,  f_{n T_B}(t) e^{-i m\omega_p t}
\end{align}
where
\begin{align}
&f_{nT_B}(t) =\nonumber\\
&\sum_{k=0}^{n-1}\Theta((k+1)T_B-t)\Theta(t-k T_B) f_{T_B}(t-k T_B)
\end{align}
is the filter function representing a pulse sequence composed of $n$ repetitions of the base sequence with total duration of $T=n T_B$. Clearly, the Fourier coefficients do not depend on $n$ and are set exclusively by the base sequence.

Within the setting described above, it is in principle possible to increase the sequence duration to an arbitrary extent, so that the spectroscopic regime can be achieved even for signals with extremely long correlation times. However, the spectroscopic formula indicates the linear $T$-scaling of the attenuation function, and consequently, the exponential decay of the coherence function $W(T)$. Therefore, there is a fine balance to be struck: on one hand $T$ must be long enough to overcome $\tau_c$, but on the other hand it must be short enough so the measured coherence still remains significant enough to yield reliable data. Depending on the overall strength of the noise and its correlation time, achieving this balance might prove to be impossible. In such a case it is certainly better to sacrifice the possibility of reaching spectroscopic regime via ``brute force,'' in favor of retaining non-zero coherence, and it becomes necessary to investigate the corrections to the spectroscopic formula,
\begin{align}
&\chispec(T) \equiv T\sum_{m>0}|\Fc{m\omega_p}|^2 S(m\omega_p)\,,\label{eq:spec_formula_def}\\
&\dchi(T) \equiv \chi(T) -\chispec(T)\,.\label{eq:dchi_def}
\end{align}
The following sections are dedicated to this task.

\section{Lorentzian spectrum}\label{sec:lorentzian}

Consider a model where the noise source (the environment) emits a signal with a spectral line at frequency $\omega_s$ (and its mirror image at $-\omega_s$ due to symmetries of Fourier transform)  yielding a spectral density of the following form:
\begin{equation}\label{eq:lorentz}
S(\omega) = \frac{S_\mathrm{L}(\omega-\omega_s)+S_\mathrm{L}(\omega+\omega_s)}{2}\,,
\end{equation}
with the line shape given by Lorentzian function
\begin{equation}\label{eq:S_L}
S_\mathrm{L}(\omega) =\int_{-\infty}^\infty \id t\, v^2 e^{-\frac{|t|}{\tau_c}} e^{-i \omega t} = \frac{ 2 v^2 \tau_c}{1+\tau_{c}^2\omega^2}\,.
\end{equation}
Here $v^2$ is the effective noise strength and $\tau_c$ is the noise correlation time. This type of line broadening is expected in, e.g. liquid or gaseous systems where the nuclear magnetic moments contributing to the emitted noise are in rapid relative motion (i.e.~the system is in the {\it motional narrowing} regime), and the central frequency $\omega_s$ can then be ascribed to precession in the external magnetic field \cite{Abragam}. A single-line Lorentzian spectrum ($\omega_s=0$) is characteristic for the Ornstein-Uhlenbeck noise \cite{Wang_RMP45}, that was shown to well describe the decoherence of electron spin qubit dipolarly coupled to a bath of electron spins \cite{deLange_Science10}. The case of Lorentzian spectrum with $\omega_s=0$ is also encountered for a qubit coupled to a source of random telegraph noise---a classical two-level fluctuator often encountered in solid state systems. While this type of noise has non-Gaussian statistics [i.e.,~it is not fully characterized by correlation function $C(|t|)$], for a weak qubit-fluctuator coupling the Gaussian approximation works quite well, see \cite{Szankowski_JPCM17} and references therein.

The attenuation function for Lorentzian spectrum can be calculated exactly using methods of contour integrals in a complex plane,
\begin{subequations}\label{eq:Lorentz_full}\begin{align}
\nonumber
\chispec(T) =&\, T \sum_{m>0}|\Fc{m\omega_p}|^2 S(m\omega_p)\\
\nonumber
=&\, v^2\, T \tau_c\Bigg[ \sum_{m>0} \frac{|\Fc{m\omega_p}|^2}{1+ \tau_c^2( m\omega_p+\omega_s)^2}\\
&\phantom{v^2\, T \tau_c}+ \sum_{m>0} \frac{|\Fc{m\omega_p}|^2}{1+ \tau_c^2( m\omega_p-\omega_s)^2}\Bigg]\,,\\
\dchi(T) =&\, -v^2\tau_c^2 \,\mathrm{Re}\Bigg[
	\left( 1 - e^{-\frac{T}{\tau_c}}e^{-i\omega_s T} \right)\nonumber\\
&\phantom{\sum_c \tau_c^2 R_{\tau_c}}\times \sum_m \frac{\Fc{m\omega_p}}{1+i\tau_c(m\omega_p - \omega_s)} \nonumber\\
\label{subEq:Lorentz_full_dchi}
&\phantom{\sum_c \tau_c^2 R_{\tau_c}}\times \sum_{m'} \frac{\Fc{m'\omega_p}^*}{1+i\tau_c(m'\omega_p - \omega_s)} 
	\ \Bigg]\,.
\end{align}\end{subequations}
The detailed derivation of this result is presented in Appendix \ref{app:lorentz}. Here we shall restrict the discussion to few key concepts which, although technical in nature, provide a valuable insight into physical origins of this particular form of the attenuation function. 

\begin{figure}[tb]
\centering
\includegraphics[width=\columnwidth]{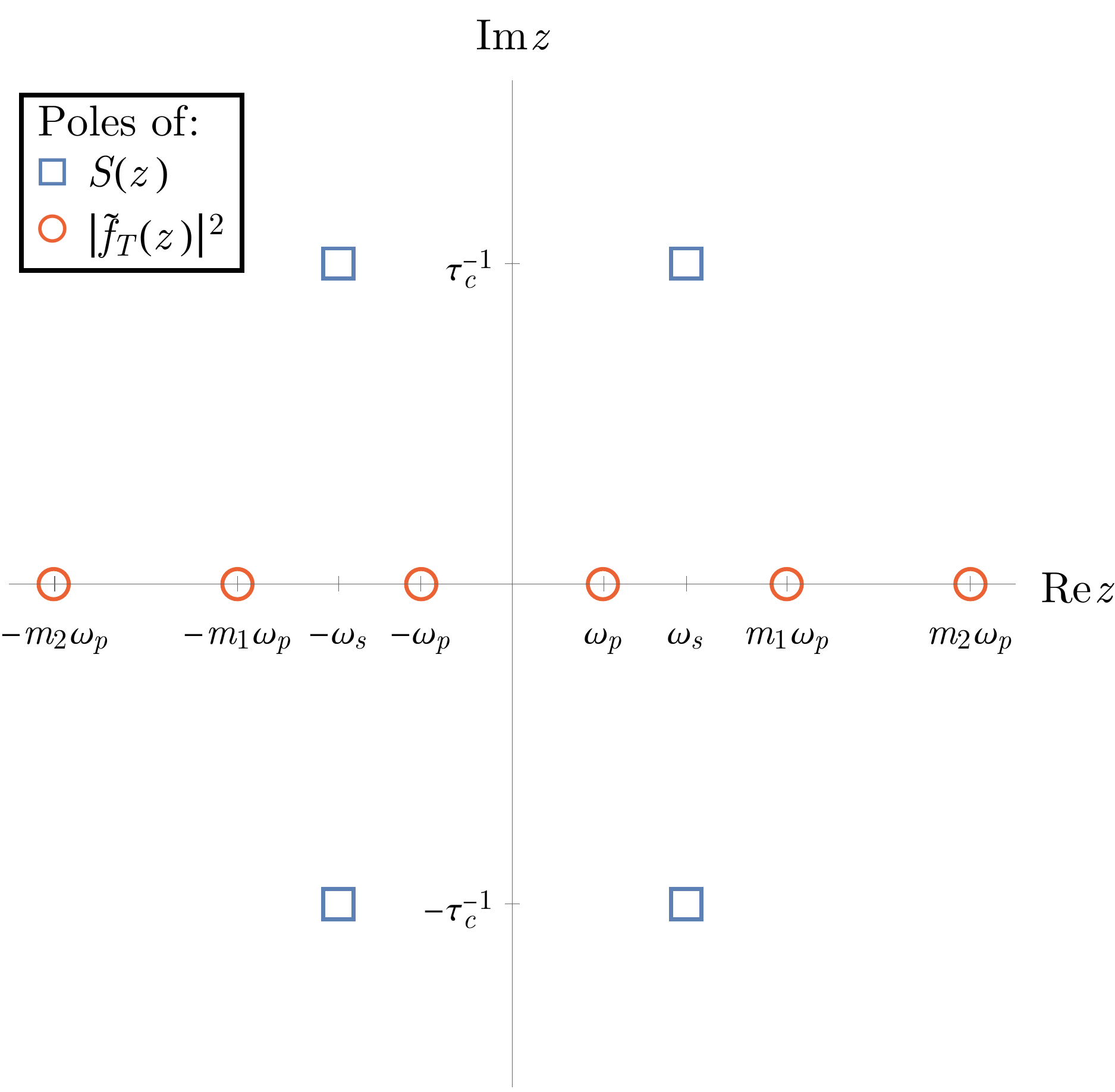}
\caption{The pole plot of Lorentzian spectrum $S(z)$ (squares) and the frequency-domain filter function $|\tilde{f}_T(z)|^2$ (circles).}\label{fig:poles}
\end{figure}
When treated as functions of a complex argument $z$, both the Lorentzian spectrum $S(z)$ and the frequency-domain filter function $|\tilde{f}_T(z)|^2$ are analytic functions in the whole complex plane except for isolated points where functions diverge---their {\it poles}. As it is depicted in Fig.~\ref{fig:poles}, the Lorentzian spectrum possess a quartet of poles, distributed symmetrically in each quadrant of the complex plane at $i\tau_c^{-1}+\omega_s,\ i\tau_c^{-1}-\omega_s,\ -i\tau_c^{-1}+\omega_s$ and $-i\tau_c^{-1}-\omega_s$. The poles of the filter function are all located on the real line, exactly at the multiples of the characteristic frequency of the pulse sequence, $m\omega_p$. (Technically speaking, $\mathrm{sinc}$ functions in \eqref{eq:freq_dom_filter} have only removable singularities, but ultimately, it is the positive or negative frequency parts $\exp[{\pm i (z-m\omega_p)T/2}]/(z-m\omega_p)$ that individually contribute to integrals.) By taking advantage of this fact, the attenuation function in the form of overlap integral \eqref{eq:attfc_freq} can be calculated utilizing the residue theorem of complex analysis. Then it is straightforward to show that the spectroscopic formula $\chispec(T)$ is a residue of poles of the ``diagonal part'' of the frequency-domain filter function \eqref{subeq:ff_diag}, while the contribution from poles of the ``off-diagonal part'', \eqref{subeq:ff_off-diag}, simply vanishes. Simultaneously, the entirety of the correction $\dchi(T)$ is composed of contribution from poles of the spectral density. This clear-cut distinction between the role of filter and spectrum poles suggest a following interpretation: the spectroscopic formula results from {\it the spectrum being filtered by the filter}, while the correction term results from a sort of role reversal, {\it the filter being filtered by the spectrum}.

With the exact form of the attenuation function \eqref{eq:Lorentz_full} in hand we may proceed to examine the conditions under which the corrections to spectroscopic formula become negligible.  For given base pulse sequence (i.e. fixed $\omega_p$ and $\Fc{m\omega_p}$) the value of $\dchi(T)$ rapidly saturates, as the only $T$-dependent part proportional to $e^{-T/\tau_c}$, goes to zero for $T\gg\tau_c$. Therefore, $\dchi(T)$ can become negligible in comparison to $\chispec(T)$ only because the linear $T$-scaling of the spectroscopic formula allows one to ramp up the sequence duration so that it overshadows the correction. The question now becomes, what is the bound on $T$ which yields a satisfactorily small relative error. In general, providing a useful answer could prove to be problematic. Indeed, in principle, the parametric dependence of a spectroscopic formula and the correction to it on both the characteristic frequency and the Fourier coefficients of the sequence could be completely different. If that would be the case, then the bounds on $T$ might vary wildly depending on which part of the spectrum we wish to probe (the choice of $\omega_p$), and the manner in which we chose to do it (the choice of $\Fc{m\omega_p}$). However, since the Lorentzian spectral density and the frequency-domain filter have similar analytical properties (most notably the presence of the power-law tail) the dependencies on $\omega_p$ and $\Fc{m\omega_p}$ of $\chispec$ and $\dchi$ are very much alike. Therefore, it is justifiable to simply ignore the $\omega_p$ and $\Fc{m\omega_p}$ dependence when comparing $\chispec$ and $\dchi$, and focus only on $T$-dependence:
\begin{equation}
\frac{|\dchi(T)|}{|\chispec(T)|} \sim \frac{\tau_c (1 - e^{-\frac{T}{\tau_c}})}{T}\xrightarrow{T\gg \tau_c} \frac{\tau_c}{T}\,.\label{eq:Lorentz_bound}
\end{equation}
According to the above formula, the relative error scales as $\tau_c/T$, and thus it can be efficiently diminished to the desired level, regardless of the choice of other pulse sequence parameters. We will encounter a very different situation for Gaussian-shaped spectrum in Sec.~\ref{sec:gauss}.

Note that relation~\eqref{eq:Lorentz_bound} is nonperturbative, i.e., due to slow convergence it is impractical to represent the correction to a spectroscopic formula as a truncated Taylor series around some {\it small} parameter. Instead, it is more appropriate to approximate it with an asymptotic series in a {\it large} parameter $T/\tau_c$. On one hand, this observation is consistent with our description of the spectroscopic approximation, where the spectroscopic regime is achieved by extending the upper limit of time integrals in Eq.~\eqref{eq:attfc_time} far beyond $\tau_c$, so that $T\gg \tau_c$. On the other hand, it suggests a significant conceptual problem in the approach where the frequency-domain filter in Eq.~\eqref{eq:attfc_freq} is approximated by a Dirac comb. As we argued previously, the Dirac delta substitution is valid when the spectrum around peaks of $|\tilde{f}_T(\omega)|^2$ can be regarded as constant. The corrections to this $0$th order term would be obtained by replacing $S(\omega)$ with its power series expansion around each peak, which would in turn lead to an undesirable Taylor series representation of \eqref{eq:Lorentz_bound}. Therefore, the Dirac comb picture of the spectroscopic approximation has only limited utility: It provides a practical quantitative description only when the corrections to the spectroscopic formula can be completely neglected, but it is not a good starting point for consideration of corrections to this description. The importance of this point becomes particularly apparent in case of spectra with finite range, e.g. Gaussian spectrum discussed in the upcoming section.

\begin{figure}[tb]
	\centering
	\includegraphics[width=\columnwidth]{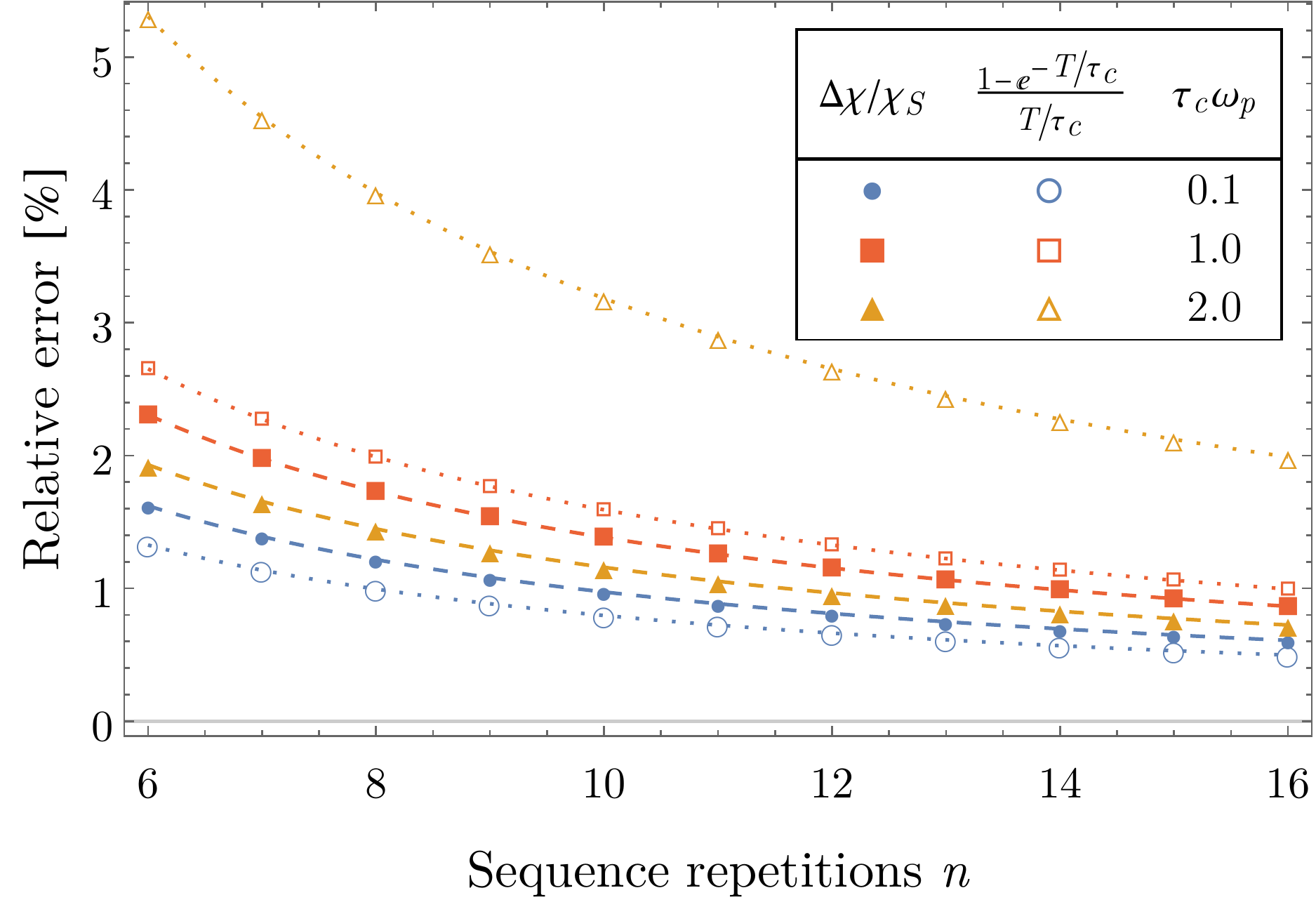}
	\caption{The comparison between $\dchi/\chispec$ (filled symbols) and the formula \eqref{eq:Lorentz_bound} (empty symbols). The spectroscopic formula $\chispec$ and the correction $\dchi$ were calculated for single line Lorentzian spectrum ($\omega_s=0$) and a base two-pulse CP sequence with characteristic frequencies set to $\tau_c\omega_p =0.1$ (circles), $1.0$ (squares), and $2.0$ (triangles), see Eqs.~\eqref{eq:CP_single_line_chispec} and \eqref{eq:CP_single_line_dchi}. The duration was manipulated by repeating the base sequences $n$ times, so that $T=n\times 2\pi/\omega_p$, where $\pi/\omega_p=\tau_p$ is the inter-pulse interval.}
	\label{fig:Lorentz_corr_vs_freq}
\end{figure}

The results of this section are illustrated with an example of a single line Lorentzian spectrum ($\omega_s=0$) probed by a qubit subjected to an $n\in\mathrm{even}$ pulse Carr-Purcell (CP) sequence defined by pulse times $t_k=(k-\frac{1}{2})\tau_p$. The characteristic frequency of CP sequence is $\omega_p = \pi/\tau_p$, its Fourier coefficients are given by $\Fc{m\omega_p}^{(\mathrm{CP})} = e^{i\frac{\pi}{2}m}\,2/(i \pi m)$ for $m\in \mathrm{odd}$ and zero otherwise, and the duration is $T=n \tau_p = n \pi/\omega_p$. (Note that the $n$ pulse CP sequence is equivalent to the $n/2$ repetitions of the base two-pulse CP sequence.) In this case the sums in Eqs.~\eqref{eq:Lorentz_full} can be carried out analytically yielding the following results
\begin{align}
\label{eq:CP_single_line_chispec}
\chispec(T) &= T\tau_c v^2\left[ 1 - \frac{2\,\tau_c\,\omega_p }{\pi}\,\mathrm{tanh}\left(\frac{\pi}{2\,\tau_c\,\omega_p}\right)\right]\,,\\
\label{eq:CP_single_line_dchi}
\dchi(T) &= -\tau_c^2 v^2(1-e^{-\frac{T}{\tau_c}})\left[1-\mathrm{sech}\!\left(\frac{\pi}{2\,\tau_c\,\omega_p}\right) \right]^2.
\end{align}
Figure~\ref{fig:Lorentz_corr_vs_freq} compares the ratio of the correction to the spectroscopic formula given by the above formula and Eq.~\eqref{eq:Lorentz_bound},  as a function of sequence duration for three values of $\tau_c\, \omega_p$.

\section{Gaussian spectrum}\label{sec:gauss}

The second case we wish to investigate is, again, a spectrum with line at $\omega_s$ (and its mirror image),
\begin{equation}\label{eq:spec_gauss}
S(\omega) = \frac{S_\mathrm{G}(\omega-\omega_s)+S_\mathrm{G}(\omega+\omega_s)}{2} \,\, ,
\end{equation}
but with the Gaussian line shape
\begin{equation}
S_\mathrm{G}(\omega) = \int_{-\infty}^\infty \id t\, v^2\frac{e^{-\frac{t^2}{2\tau_c^2}}}{\sqrt{2\pi}\tau_c}e^{-i \omega t}=v^2 e^{-\frac{1}{2}\tau_c^2\omega^2}\,.
\end{equation}
This form of line broadening is found, e.g. in magnetic noise emitted by immobile nuclear magnetic moments of crystal lattice \cite{Abragam}, and thus it is an expected line-shape when a qubit such as a nitrogen-vacancy center is used to sense a group of nuclear spins from a nearby protein attached to the surface of a diamond nano-crystal \cite{Lovchinsky_Science16}. Since the Lorentzian spectrum is characterized by its long tail, the feature obviously lacking in Gaussian spectrum, one should expect some fundamental differences between their respective correction to the spectroscopic formula.

The first major difference is that the contour integrals method cannot be employed to calculate the attenuation function for the Gaussian spectrum. In short, since $S_\mathrm{G}(z=i R) \propto e^{R^2} \xrightarrow{R\to\infty}\infty$, it is impossible to devise an appropriate closed contour which would yield vanishing integrals when its radius is stretched to infinity (compare with Appendix \ref{app:lorentz}). Unfortunately, this implies that the clear-cut distinction between the roles of $S(\omega)$ and $|\tilde{f}_T(\omega)|^2$ we were able to find for Lorentzian spectrum cannot be invoked here.

The simplest approach to calculating the exact form of the attenuation function for the Gaussian spectrum we where able to identify is to instead remain in time-domain [i.e., to use Eq.~\eqref{eq:attfc_time}] and to utilize the {\it error function} defined as
\begin{equation}
\erf(z) = \frac{2}{\sqrt \pi}\int_0^z \id t\, e^{-t^2}\,.
\end{equation}
Note that for complex $z$ it is implied that the integration is performed over a curve in a complex plane with fixed endpoint at $0$ and $z$. Since $e^{-z^2}$ is an analytical function and it does not posses any poles, the path of integration can be chosen arbitrarily, as the result is independent of its course.

The resultant expression for attenuation function have a familiar structure of ``diagonal'' and ``off-diagonal'' sums over multiples of characteristic frequency,
\begin{widetext}\begin{align}
\nonumber
\chi(T) =&\ v^2\sum_{m>0}|\Fc{m\omega_p}|^2 \bigg[\g(\omega_s+m\omega_p) + \g(\omega_s-m\omega_p) \bigg]\\
\label{eq:attfc_gauss}
&+v^2\!\!\!\sum_{m_1\neq -m_2}\frac{\Fc{m_1\omega_p}\Fc{m_2\omega_p}}{m_1\omega_p+m_2\omega_p}
	\bigg[ \f(\omega_s-m_1 \omega_p) - \f(\omega_s+m_2 \omega_p) \bigg]\,,
\end{align}\end{widetext}
where the auxiliary $\f$ and $\g$ are given in terms of error functions:
\begin{subequations}\begin{align}
\label{subeq:aux:f_T}
\f(\omega) =&\  \frac{1}{2}e^{-\frac{\tau_c^2\omega^2}{2}}\mathrm{Im}\left\{ \erf\left(\frac{T}{\sqrt 2 \tau_c} - i \frac{\tau_c\, \omega}{\sqrt 2}\right)\right\}\\[.1cm]
\label{subeq:aux:f_0}
&+\frac{1}{2i}e^{-\frac{\tau_c^2\omega^2}{2}}\erf\left(i \frac{\tau_c\, \omega}{\sqrt 2}\right)\,,
\end{align}\end{subequations}
\begin{subequations}\begin{align}
\nonumber
\g(\omega) =&\  \frac{\tau_c}{\sqrt{ 2\pi}}e^{-\frac{T^2}{2\tau_c^2}}\cos \omega T+ \\
\label{subeq:aux:g_T}
&\mathrm{Re}\left\{ (T - i \tau_c^2 \omega)\,\erf\left(\frac{T}{\sqrt 2 \tau_c} - i \frac{\tau_c\, \omega}{\sqrt 2}\right)\right\}\\[.1cm]
\label{subeq:aux:g_0}
&\ -\frac{\tau_c}{\sqrt{2\pi}}
	+\frac{1}{2i}\tau_c\, \omega\,e^{-\frac{\tau_c^2\omega^2}{2}}\erf\left(i\frac{\tau_c\,\omega}{\sqrt 2}\right)\,.
\end{align}\end{subequations}
Although exact, the above expressions are not transparent enough to allow for clear interpretation and convincing estimation of spectroscopic regime conditions. Hence, we are forced to simplify them with an application of carefully tailored approximations, which we proceed to discuss below.

\begin{figure}[tb]
\centering
\includegraphics[width=\columnwidth]{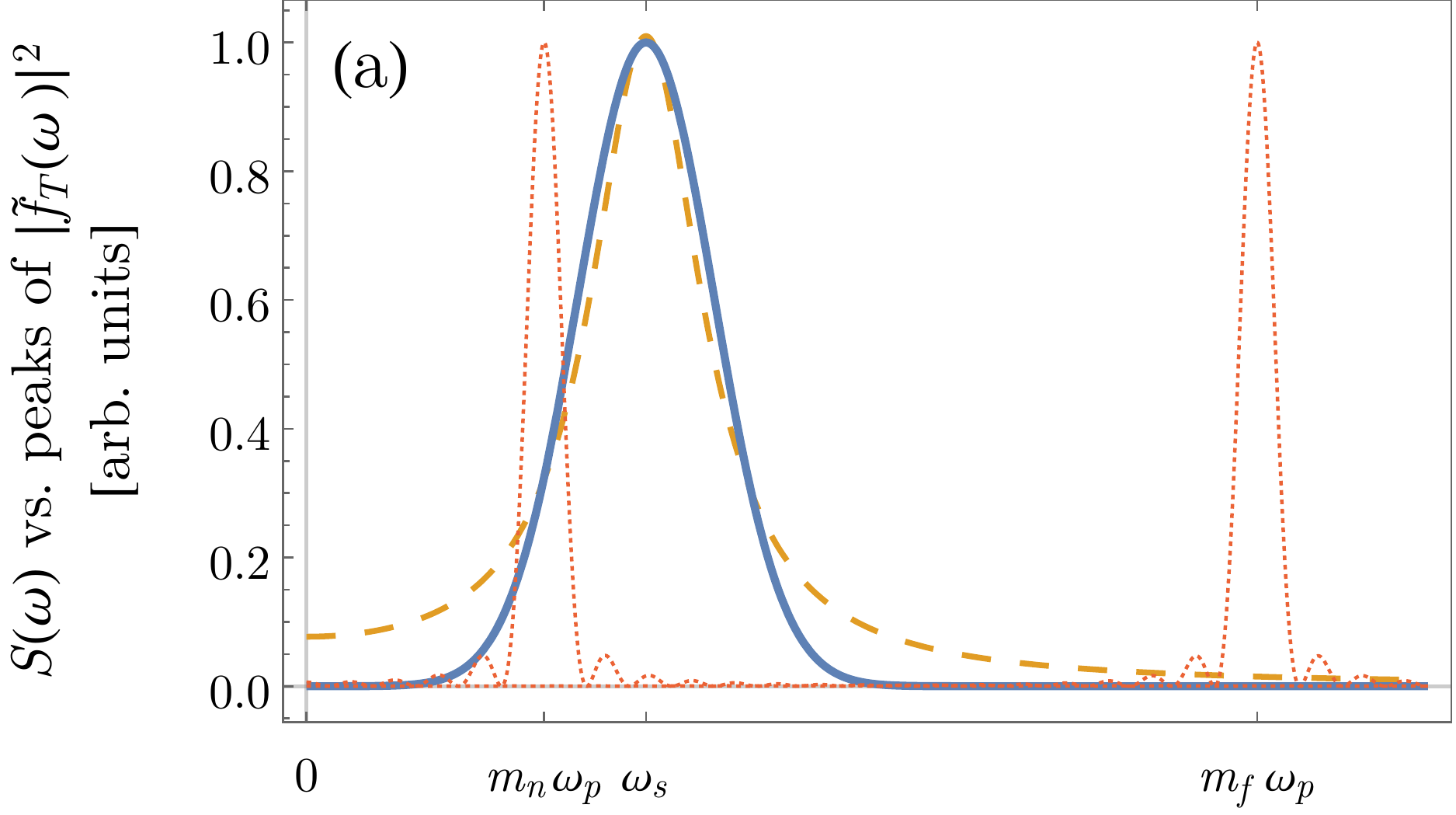}
\includegraphics[width=\columnwidth]{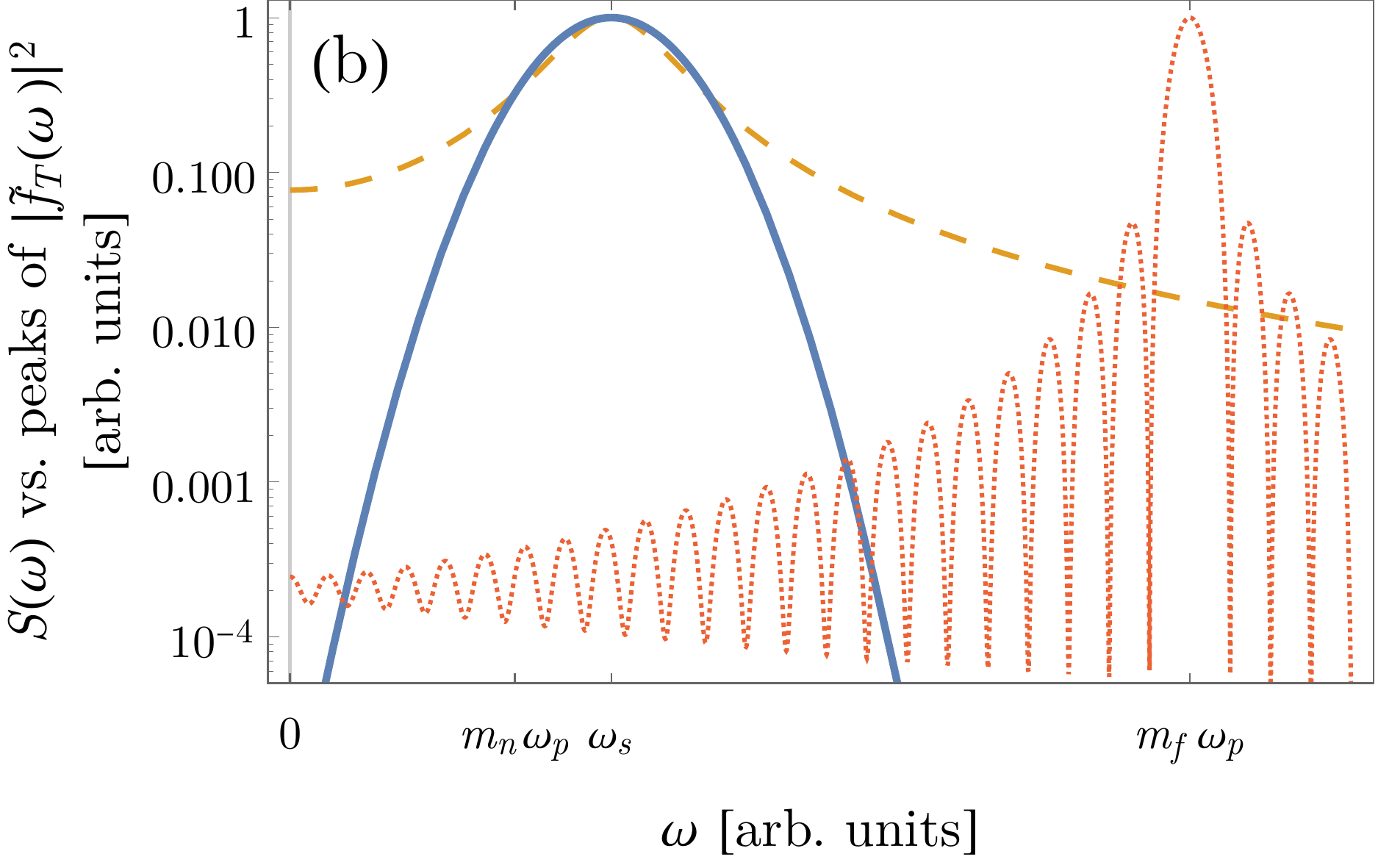}
\caption{The schematic diagram illustrating the concept of {\it near} and {\it far} frequency-domain filter peaks. The spectral line of Gaussian spectrum is centered around $\omega_s$ (solid blue line). In panel (a), the two $\mathrm{sinc}^2(T(\omega-m\omega_p)/2)$ shaped peaks depicted here (dotted red lines) are centered at frequencies $m_n\omega_p$ and $m_f \omega_p$. One of them clearly overlaps with the spectrum, so that $\tau_c|m_n \omega_p - \omega_s|\lesssim 1$, and hence it is considered as a near peak. The other one is positioned away from the center of the spectral line, so that $\tau_c|m_f\omega_p-\omega_s|\gg 1$, and in consequence is counted among far peaks. Panel (b) shows the overlap between the far peak and the Gaussian spectral line in the log scale. The distinction between far and near peaks is impossible to establish for long-tailed spectra, such as Lorentzian given by Eq.~\eqref{eq:lorentz} and \eqref{eq:S_L}, and depicted here in orange, dashed line. (The correlation time is the same for both spectra.) The spectra and filter peaks have been normalized here to be of the same height.}
\label{fig:near_and_far_peak}
\end{figure}

First, let us consider in brief the frequency-domain overlap integral form of attenuation function \eqref{eq:attfc_freq}. Even though, not useful for the purpose of concrete calculations, this form provides helpful insight into structure of the expression. For example, by comparing terms proportional to a given combinations of Fourier coefficients in \eqref{eq:attfc_freq} and \eqref{eq:attfc_gauss}, we can see that the diagonal part of $\chi$ results from the overlap with peaks of \eqref{subeq:ff_diag}, while the off-diagonal part from the overlap with peaks of \eqref{subeq:ff_off-diag}. Carrying this peaks-overlapping-with-spectrum picture to its logical conclusion, we can divide terms in $\chi$ into two groups [see panel (a) of Fig.~\ref{fig:near_and_far_peak}]: (i) terms resultant from the overlap with peaks centered at frequencies $m\omega_p$ that lie {\it far} beyond the range of the spectrum, i.e., $\tau_c|\omega_s - m \omega_p|\gg 1$, (ii) terms resultant from peaks lying {\it near} the spectrum, so that $\tau_c|\omega_s - m\omega_p|\lesssim 1$. Note that such a divide between filter peaks is impossible for Lorentzian spectrum. The power-law decay of its tails is not abrupt enough to facilitate a sharply defined border between far and near frequencies. Of course, in the case of Gaussian spectrum, such a border exists, virtually by the very definition of finite-ranged spectral density. Having established this hierarchy of terms, the need to account for overlap with near peaks is certainly indisputable. On the other hand, it might seem reasonable to simply neglect whole contributions from far peaks, since the Gaussian spectrum decays very rapidly beyond its range. It is essential to recognize that this would be incorrect, because filter peaks possess long tails, see panel (b) of Fig.~\ref{fig:near_and_far_peak}. Therefore, even for far peaks, the cumulative contribution from the overlap of their tails with the spectrum can be substantial, and should be included in some form. This is especially crucial when no peaks of the filter can be considered near---the contribution from the overlap of the peak tail with the spectrum is then the {\it only} nonvanishing term in the attenuation function.

Next, the minimal requirement for robust noise spectroscopy is the sequence duration to be much longer than the correlation time. Then, the matter of real interest is the manner in which the spectroscopic formula dominates over corrections to it. Therefore, it is justifiable to examine the exact formula for the attenuation function in the limit $T \gg \tau_c$. This is in agreement with the comments on correction approximation from a previous section, which suggested the use of an asymptotic expansion for $\chi(T/\tau_c)$. 

Both the limit of long duration, and of far peaks can be treated within the same framework of the asymptotic series expansion. More precisely, since the difficulty with the interpretation of result \eqref{eq:attfc_gauss} is directly tied to error functions, it is enough to approximate them with their asymptotic series:
\begin{align}
&\mathrm{Re}\,{z} > 0\,:\ \erf(z) \stackrel{|z|\gg 1}{\approx}\!\! 1+  \sum_{k=1}^\infty \frac{(-1)^k\,\Gamma(k-\frac{1}{2})}{\pi} \frac{e^{-z^2}}{z^{2k-1}}\nonumber\\
	&\phantom{\mathrm{Re}\,{z} > 0\,:\ \erf(z)}\approx 1 - \frac{1}{\sqrt \pi} \frac{e^{-z^2}}{z}+\frac{1}{2\sqrt\pi}\frac{e^{-z^2}}{z^3}\, ,\\[.3cm]
&\mathrm{Re}\,{z} = 0\, :\  \erf(z=iy) \stackrel{|y|\gg 1}{\approx} -\frac{1}{\sqrt \pi} \frac{e^{y^2}}{iy}+\frac{1}{2\sqrt\pi}\frac{e^{y^2}}{(i y)^3}.
\end{align}
The first series is relevant for $T$-dependent parts of $\f$ and $\g$ (Eqs.~\eqref{subeq:aux:f_T} and \eqref{subeq:aux:g_T}), both for far and near peak terms. The second series can be applied only to $T$-independent parts of auxiliary functions (Eqs.~\eqref{subeq:aux:f_0} and \eqref{subeq:aux:g_0}), and only to far peak terms where $\erf$ arguments satisfy $\tau_c|\omega|\gg 1$. Expanding an appropriate expressions to the lowest order recovers the spectroscopic formula $\chispec(T)\propto T$ and the correction term, which we split into three parts: the $T$-dependent $\dchi_T$ and far and near peak $T$-independent $\dchi_{0,\mathrm{far}/\mathrm{near}}$,
\begin{widetext}\begin{align}
&\chi(T) \stackrel{\frac{T}{\tau_c}\gg 1}{\approx}\ \chispec + \dchi_T+\dchi_{0,\mathrm{far}} + \dchi_{0,\mathrm{near}}\,,\\[.5cm]
&\chispec = T \sum_{m>0}|\Fc{m\omega_p}|^2 S(m\omega_p) = Tv^2\sum_{m>0}|\Fc{m\omega_p}|^2
	\frac{e^{-\frac{\tau_c^2(\omega_s-m\omega_p)^2}{2}} + e^{-\frac{\tau_c^2(\omega_s+m\omega_p)^2}{2}}}{2}\,,\\[.3cm]
\label{eq:dchi_T}
&\dchi_T =\frac{v^2\tau_c}{\sqrt{2\pi}}\, e^{-\frac{T^2}{2\tau_c^2}}\,\mathrm{Re}\Bigg[\ e^{i\omega_s T}
	\sum_m \frac{\Fc{m\omega_p}}{\left(\frac{T}{\tau_c}+i \tau_c(m\omega_p-\omega_s) \right)^2}
	\sum_{m'} \frac{\Fc{m'\omega_p}}{\left(\frac{T}{\tau_c}-i \tau_c(m'\omega_p+\omega_s) \right)^2}\ \Bigg]
	\sim v^2\tau_c^2 \left(\frac{\tau_c}{T}\right)^4\frac{e^{-\frac{T^2}{2\tau_c^2}}}{\sqrt{2\pi}\tau_c}\,,\\[.3cm]
\label{eq:dchi_far}
&\dchi_{0,\mathrm{far}} \approx \frac{v^2}{\sqrt{2\pi}\tau_c}
	\sum_{m_f\in \mathrm{far}} \frac{\Fc{m_f\omega_p}}{\omega_s+m_f\omega_p}
	\sum_{m_f'\in \mathrm{far}} \frac{\Fc{m_f'\omega_p}}{\omega_s-m_f'\omega_p}\,,\\[.3cm]
\nonumber
&\dchi_{0,\mathrm{near}} =
	 v^2\sum_{\substack{m_n\in\mathrm{near} \\ m_n>0 }}|\Fc{m\omega_p}|^2\
	 \left[ \g_0(\omega_s-m_n\omega_p)+\g_0(\omega_s+m_n\omega_p)\right]\\
&\phantom{\dchi_{0,\mathrm{near}} =}
	+v^2\sum_{\substack{m_1,m_2 \in \mathrm{near} \\ m_1 \neq -m_2}}\frac{\Fc{m_1\omega_p}\Fc{m_2\omega_p}}{m_1\omega_p+m_2\omega_p}
	 \left[\f_0(\omega_s-m_1\omega_p)-\f_0(\omega_s+m_2\omega_p)\right]\,.
\end{align}\end{widetext}
Here $\f_0$ and $\g_0$ are the $T$-independent parts of auxiliary functions, given by Eq.~\eqref{subeq:aux:f_0} and \eqref{subeq:aux:g_0} respectively. We also introduced a shorthand notation $m\in\mathrm{far}$ which indicates that the sum runs over only such $m$-s that satisfy $|m|\omega_p-\omega_s \gg\tau_c^{-1}$, and $m\in\mathrm{near}$ to indicate sums over all other $m$-s.
\begin{figure}[tb]
\centering
\includegraphics[width=\columnwidth]{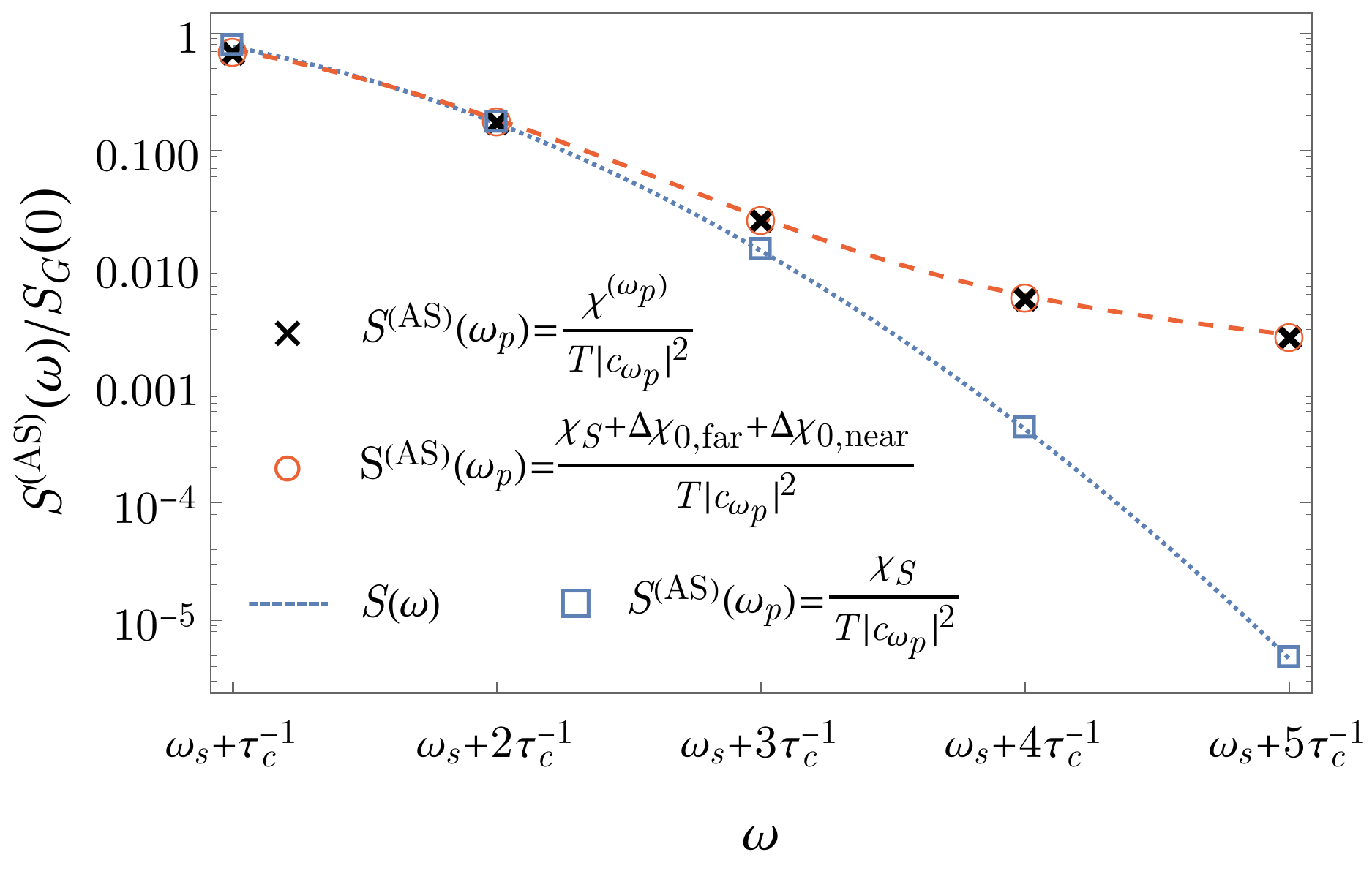}
\caption{An example of Gaussian spectrum reconstruction using the simplest version of \as{} method where the cut-off is set to $m_c=1$. The recovered value of spectral density is given by $S^{(\mathrm{AS})}(\omega_p) = \chi^{(\omega_p)}(T)/T|\Fc{\omega_p}|^2$ (crosses). The measured attenuation functions $\chi^{(\omega_p)}(T)$ were simulated via numerical integration of Eq.~\eqref{eq:attfc} using correlation function $C(|t|)=(\sqrt{2\pi}\tau_c)^{-1}e^{-t^2/2\tau_c^2}\cos\omega_s t$ ($\omega_s=\pi\tau_c^{-1}$) and the time-domain filter functions of a $n=16$ pulse CP sequences with  characteristic frequencies ranging from $\omega_p=\omega_s+\tau_c^{-1}$ to $\omega_p=\omega_s+5\tau_c^{-1}$. The number of pulses in each sequence was always the same, hence there was no attempt to ramp up the duration in order to compensate for rapid decay of the spectral density. The figure shows on the log scale the spectra $S^{(\mathrm{AS})}$ reconstructed with the exact $\chi^{(\omega_p)}$ (crosses) and the approximate $\chi^{(\omega_p)}\approx \chispec+\dchi_{0,\mathrm{near}}+\dchi_{0,\mathrm{far}}$ (circles), and compares them with the real course of the spectral density $S(\omega)$ (dotted line). The latter is in fact approximated very well by the spectroscopic formula, $S^{(\mathrm{AS})}(\omega_p)\approx\chi_{S}/T|c_{\omega_p}|^2\approx S(\omega_p)$ (squares). Note how the corrections $\Delta\chi_{0,\mathrm{far}}$, $\Delta\chi_{0,\mathrm{near}}$ dominate over $\chispec$, which then leads to erroneous attribution of the long-tailed behavior to the reconstructed spectrum.}
\label{fig:chi_vs_spec_formula}
\end{figure}

For $T\gg \tau_c$, as seen in Eq.~\eqref{eq:dchi_T}, the extremely rapid decay of the absolute value of $\dchi_T$ renders it completely negligible, as it can be made arbitrarily small (e.g. smaller then the expected measurement errors) with minimal increments of the duration. Therefore, analogous to the case of Lorentzian spectrum, also for Gaussian spectrum the mechanism behind spectroscopic approximation is the linear $T$-scaling of the spectroscopic formula $\chispec$ dominating over the $T$-independent $\dchi_{0,\mathrm{near}}$ and $\dchi_{0,\mathrm{far}}$. However, in contrast to Lorentzian, the differences in analytical properties of Gaussian spectrum and the filter function lead to significant differences in parametric dependence of $\chispec$ and the correction on $\omega_p$ and $\Fc{m\omega_p}$. In certain circumstances, this feature of finite range spectra puts a substantial constraint on applicability of spectroscopic approximation. For example, consider a case when CP sequence is used, and characteristic frequency is set to be significantly greater than central frequency of the spectral line, $\omega_p-\omega_s\gg \tau_c^{-1}$. Additionally, the number of pulses $n$ is set high enough so that the duration of the sequence, $T= n \pi/\omega_p$, is much longer than $\tau_c$ and $\dchi_T\approx 0$. In these settings, the filter is exclusively composed of far peaks and the correction is given by $\dchi\approx\dchi_{0,\mathrm{far}}(\omega_p)\sim (\omega_s/\omega_p)^4$. The spectroscopic formula can then be approximated by retaining only the first term of series, $\chispec(\omega_p) \sim T \exp\left[-\tau_c^2(\omega_s-\omega_p)^2/2\right]$; obviously it decays much faster than $\dchi(\omega_p)$ with increasing $\omega_p$. Therefore, the cost of maintaining a constant level of relative error, paid in the number of additional pulses, ramps up super-polynomially when the sequence probes the edges of spectral line. This effect might lead to some difficulties with spectrum reconstruction for unaware observer. Unless an a priori knowledge allows one to anticipate the need for such an extreme ramp up of the duration (provided that this option in even within the technical limitations of the experiment), the correction eventually overtakes the finite range spectrum, appearing as if the long tail is an actual feature of spectral density. Figure~\ref{fig:chi_vs_spec_formula} depicts an example of a scenario where such erroneous attribution would take place. In contrast, in the case of Lorentzian spectrum reconstruction, pitfalls like this are never encountered. Indeed, when the pulse sequence probes frequencies distant from the line center, $\omega_p\gg\omega_s$, the correction behaves as $\sim (1/\tau_c\omega_p)^4$---a generic feature for spectra with power-law tails (see Appendix~\ref{app:correction_est}). This means that the decay of the correction is actually faster than that of the spectroscopic formula, which depends on characteristic frequency as $\sim (1/\tau_c\omega_p)^2$. Therefore, in this case, when $\omega_p$ is increased, the relative error improves even without the need for application of additional pulses. However, if the exponent of the power-law spectrum is greater, i.e. $S(\omega)\to (1/\omega)^\beta$ with $\beta> 4$, the relative error might become larger with the increase of $\omega_p$, unless it is compensated for by the ramped up duration. Of course, this effect will never be as oppressive as in the case of finite-ranged spectral densities.

It should be reiterated that the difficulties described above are of concern when one intends to reconstruct the tails of the spectral density. However, since the spectrum in the tail region is  naturally much smaller than in vicinity of the spectral line center, the data required for tail reconstruction must be acquired with large enough accuracy to allow for observation of relatively small decay of the coherence. While recent experiments (see e.g.~\cite{Lovchinsky_Science16}) have not reached such accuracy yet, the ultimate experimental goal is to reach much higher precision, and then knowing about pitfalls such as the one described above becomes crucial.

\section{The modified data acquisition scheme for spectrum reconstruction method}\label{sec:fix}

The discussion of the corrections to the spectroscopic formula for Gaussian spectrum revealed a serious drawback of standard approach to noise spectroscopy. The data required for the reconstruction of the spectral line shape, with limited prior knowledge of its character, has to be acquired with a large spread of pulse sequence characteristic frequencies $\omega_p$. For spectra with finite range, as the scanning frequency of the sequences strays from the center of the line, the contribution from the long tailed corrections start to dominate over rapidly decaying spectral density. In contrast to long-tailed spectra, this decay cannot be efficiently counteracted by the linear amplification with the sequence duration $T$ of the spectroscopic formula. For this reason, it can become difficult to reliably distinguish whether one records the actual long tail of the spectrum, or merely an artifact of the method. Fortunately, our previous considerations also suggest a workaround for this conundrum.

The common feature of both the Lorentzian and the Gaussian spectrum is the rapid decay of the $T$-dependent part of the correction to the spectroscopic formula. In fact, a rough estimate for an arbitrary spectrum (see, Appendix~\ref{app:correction_est}) shows that this decay is not a coincidence, and that it is at least as fast as $C(|T|)$---the correlation function of the noise at $T$. It follows, that as the duration becomes long in comparison to $\tau_c$, the only $T$-dependence in the attenuation function that remains is the linear scaling of the spectroscopic formula. Therefore, the contribution from long tailed correction can be circumvented by performing a linear fit $F(T)= a \,T + b$ to data points gathered in a series of measurements with fixed $\omega_p$ and $\Fc{m\omega_p}$, and progressively longer duration. The slope of the fit to the data points for which the abscissa satisfies $T\gg\tau_c$, is simply
\begin{equation}
a=\sum_{m>0}|\Fc{m\omega_p}|^2 S(m\omega_p) = \frac{\chispec(T)}{T}\,,
\end{equation}
while the intercept is 
\begin{equation}
b=\lim_{T\to\infty}\dchi\,,
\end{equation}
i.e. the duration independent part of the correction. The intercept $b$ can then be discarded and the slope $a$ fed as an input to the standard \as{} method for the spectrum reconstruction. 

\begin{figure}[tb]
\centering
\includegraphics[width=\columnwidth]{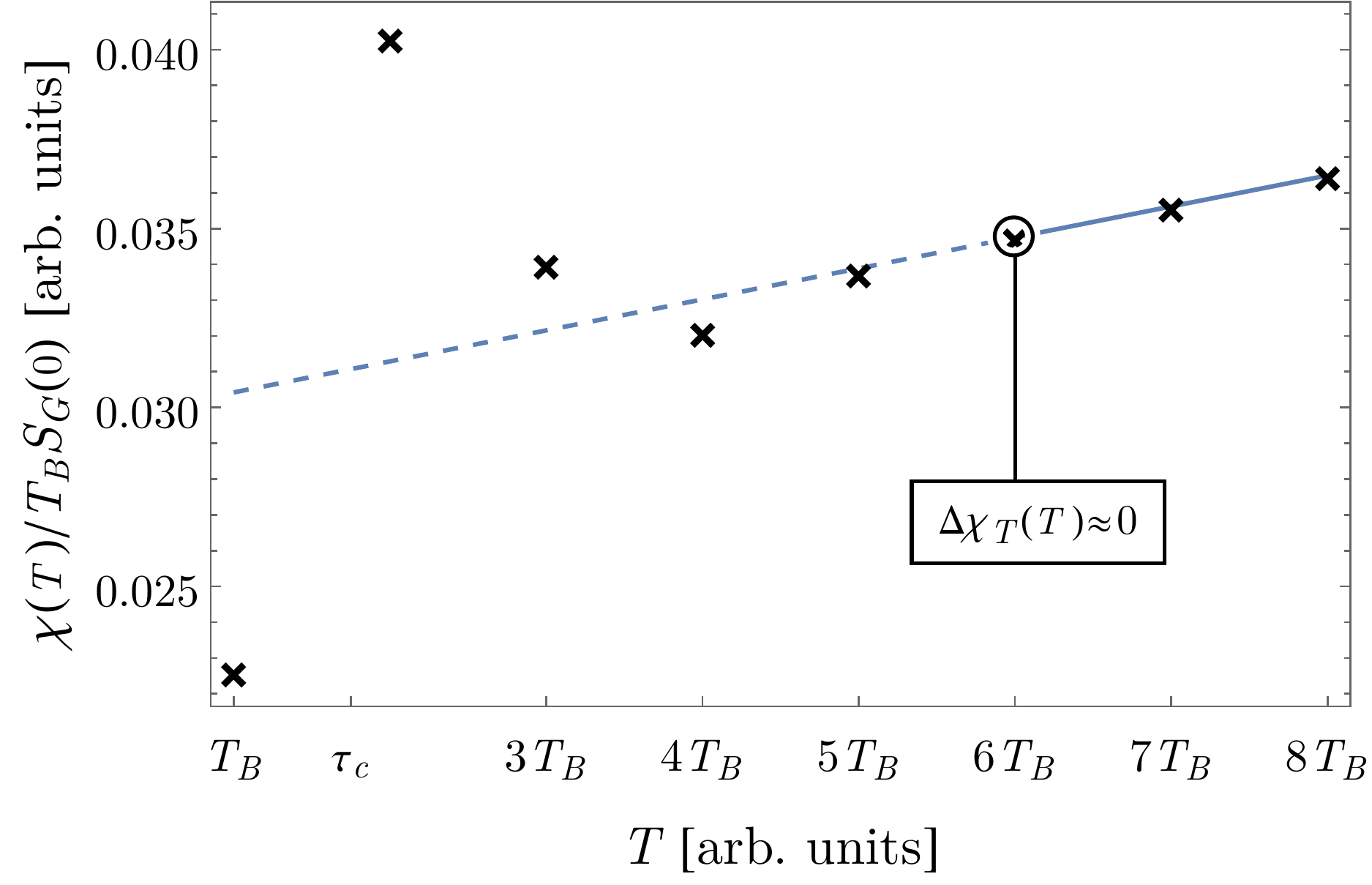}
\caption{An example of implementation of the modified data acquisition scheme. The data points plotted on the figure (crosses) represent the values of the attenuation function gathered in a series of numerically simulated measurements with progressively longer duration. Here $T_B$ is the duration of 2-pulse CP sequence employed in the measurement of the first datum. The $n$th data point was obtained with a sequence composed of $n$ repetitions of the first sequence, so that $\omega_p$ and $\Fc{m\omega_p}$ were fixed, while the duration was extended to $n T_B$. The figure shows that for durations comparable with $\tau_c$, when the $T$-dependent correction is still significant, the data points are scattered around the linear trend. Only after a number of duration increments, $\dchi_T$ becomes small enough, so that the points start to fall on a line. Hence, from the course of the plot we can determine that $T\approx 6T_B$ is a minimal duration that satisfies the condition $T\gg\tau_c$. The following values of parameters where chosen for the simulations: $\omega_p=1,\omega_s=0.7\omega_p,T_B=2\pi/\omega_p,\tau_c=3.5\pi/\omega_p$ and the noise spectrum was given by Eq.~\eqref{eq:spec_gauss}. }\label{fig:fit}
\end{figure}

The application of the above scheme solves two major problems of standard noise spectroscopy. (i) The use of the slopes $a$ of the linear fit as an input for \as{} reconstruction method completely removes the contribution from the corrections to the spectroscopic formula. This obviously results in an improvement of the accuracy of the noise spectroscopy, but more importantly, it eliminates the possibility of accidental attribution of incorrect long tail behavior to the reconstructed spectral densities. (ii) The second is a more subtle, but nevertheless, important problem. As it was discussed multiple times, a successful implementation of noise spectroscopy hinges on having the ability to set the duration so that $T\gg\tau_c$. More precisely, what is required is that the $T$-dependent correction to the spectroscopic formula, $\dchi_T$, is negligible, for which the condition $T\gg\tau_c$ is sufficient (but not necessary). However, since the noise spectrum is unknown until it is reconstructed, it is unreasonable to expect one to know in advance the value of $\tau_c$. If so, then how one could anticipate what value of $T$ is long enough to enable a reliable reconstruction? Fortunately, the problem of ascertaining whether the condition $T\gg\tau_c$ is satisfied (or more generally, whether $\dchi_T\approx 0$) can be solved by inspecting the data points used for the linear fit. Ideally, the values of the attenuation function, gathered in a measurement series would form a pattern similar to one shown in Fig.~\ref{fig:fit}. The scattering of the initial points around the asymptotic linear trend signifies that $\dchi_T$ is not yet negligible. This means that the durations corresponding to those points are comparable with $\tau_c$ (which also provides an estimate for noise correlation time, even before the spectrum is fully reconstructed). Therefore, the condition $T\gg\tau_c$ is satisfied for values of $T$ greater than the duration beyond which the data points start to fall on a line, and the slope of this line is the suitable input for \as{} method. If all the data points fit to linear function, the slope can be obtained readily, at a rather low cost of lost opportunity for independent $\tau_c$ estimation. More attention should be drawn to the eventuality when the data points are not fit well with a linear function. Encountering such a case should be read as a signal that the gathered data is not suited for the purpose of spectrum reconstruction, because either the durations used in the measurement series did not satisfy $T\gg\tau_c$, or some of the fundamental assumptions of the theory are not met (e.g. the noise is not stationary, and hence, the spectral density is not defined). Obviously, such an ``early warning system'' is of a great value, as it allows to avoid mistakes which might have been missed if the \as{} method would be applied uncritically.

\section{The error estimation for modified \as{} method}\label{sec:cut-off_error}

As it was touched upon in section \ref{sec:statement}, the shape of the spectral line can be reconstructed with \ASM{}, which takes as an input the measured attenuation functions acquired in a series of experiments performed with a properly chosen variety of pulse sequences. The first basic requirement for the method to work---achieving the spectroscopic regime---can be effectively satisfied by employing the linear fit scheme described in Sec.~\ref{sec:fix}. Then, the inputs expected by the reconstruction method in a form of normalized attenuation functions $\chi^{(j)}(T_j)/T_j$ are replaced by slopes $a_j$ found for each pulse sequence. The vector relation between inputs and unknown spectrum values, \eqref{eq:AS_eq_sys}, is now rewritten as
\begin{equation}
a_j = \sum_{m} U_{j,m}S_m\,.
\end{equation}
As before, $S_m$ is the vector of unknown values of spectrum at frequencies picked out by pulse sequences, and $\mathbf U$ is the matrix of known Fourier coefficients. In order to invert this relation, the cut-off for dimensions of $\mathbf U$ has to be assumed, which is equivalent to approximation where the infinite sum over frequencies in a slope $a_j$ (or in the spectroscopic formula) is truncated at some $m_c$:
\begin{align}
\nonumber
&\left(\text{$\mathbf{U}$ is of finite dimension}\right) \Leftrightarrow \\
&a_j = \sum_{m>0}|\Fc{m\omega_j}^{(j)}|^2 S(m\omega_j) \approx \!\!\!\!\sum_{\substack{m>0 \\m\omega_j\leqslant m_c\omega_1}}\!\!\!\!\!|\Fc{m\omega_j}^{(j)}|^2S(m\omega_j)\,.
\end{align}
Therefore, the second requirement for successful implementation of the method is to introduce an error to its results with the cut-off $m_c$. Here, our goal is to estimate this error.

\begin{figure}[tb]
\centering
\includegraphics[width=\columnwidth]{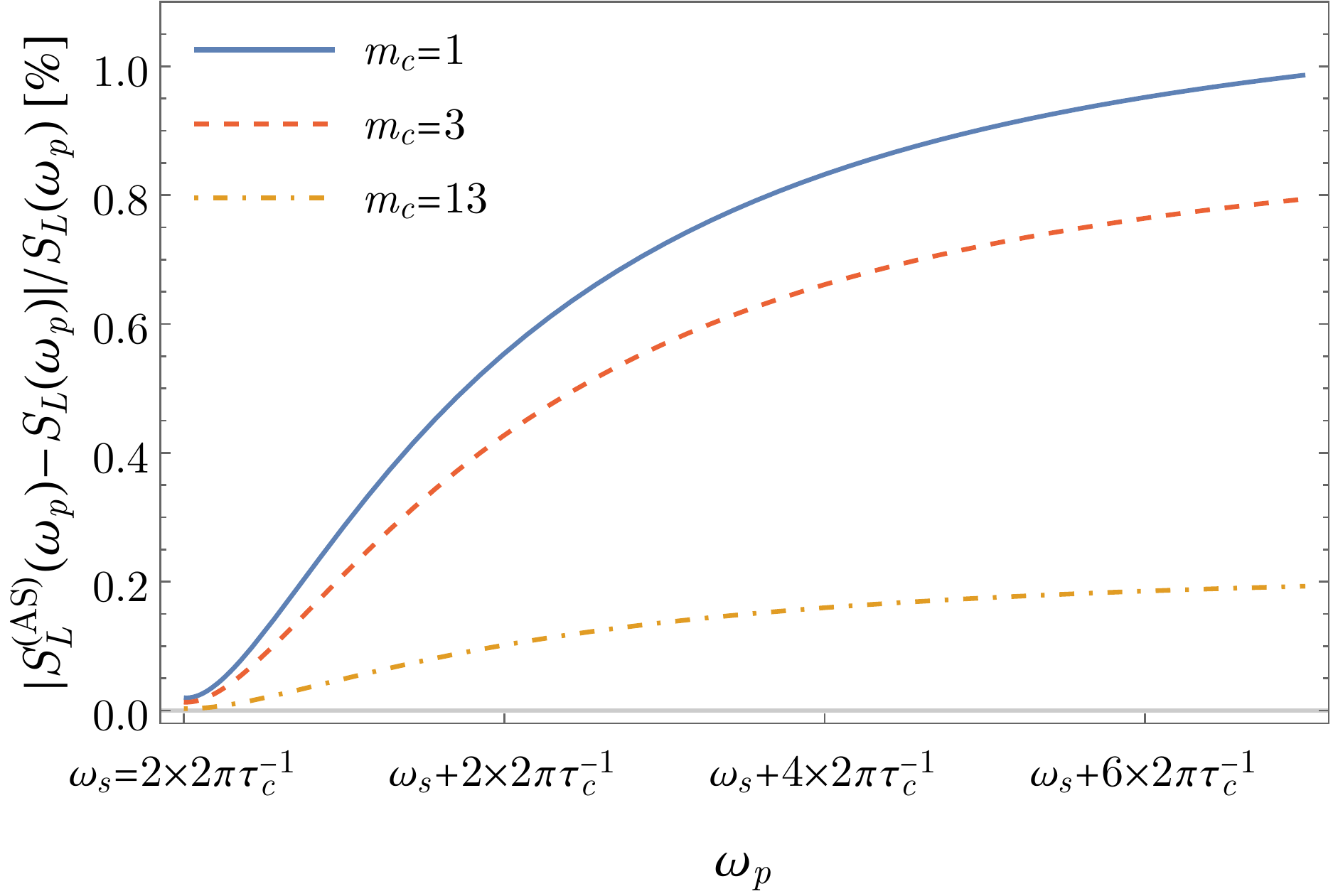}
\caption{The bound for the relative error of \as{} method \eqref{eq:AS_error_bound} for the reconstruction of Lorentzian spectrum \eqref{eq:lorentz} with three choices of cut-off $m_c$.}\label{fig:AS_error}
\end{figure}
We focus on the case where the \ASM{} is implemented with CP sequences exclusively. The characteristic frequency of the first sequence is chosen arbitrarily as $\omega_1 = \omega_p$. The remaining sequences have their frequencies set to subsequent harmonics of the first sequence: $\omega_2 = 3\omega_p,\ldots,\omega_j = (2j-1)\omega_p,\ldots, \omega_{j_c} = (2j_c-1)\omega_p = m_c\omega_p$, so that the totality of $(m_c+1)/2$ slopes $a_j$ can be generated. When we assume that the infinite series in each $a_j$ was truncated beyond $m\omega_j \leqslant m_c \omega_1 = m_c\omega_p$, the elements of matrix $\mathbf{U}$ can be written in a compact way as
\begin{equation}
U_{j,m} = \sum_{n=1}^\infty \frac{4}{\pi^2}\frac{1}{n^2}\delta_{m,(2j-1)n}\,,
\end{equation}
where index $j$ runs from $1$ to $j_c =(m_c+1)/2$, while $m=1,3,\ldots,m_c$. The value of spectrum at $\omega_p$ can be recovered by applying the inverse of $\mathbf{U}$
\begin{equation}\label{eq:SASS}
S^{(\mathrm{AS})}(\omega_p) = \sum_{j=1}^{j_c} U_{1,j}^{-1} a_{j}\,.
\end{equation}
The superscript $(\mathrm{AS})$ reminds us that it is an approximation, and this value would equal real $S(\omega_p)$ only if the assumed cut-off was in effect. The inversion of $\mathbf{U}$ can be done analytically, and the relevant matrix elements are given by the following
\begin{equation}
U_{1,j}^{-1} = \frac{\pi^2}{4}\left\{ \begin{array}{lcl}
	-\frac{1}{(2j-1)^2} &\mathrm{if}& \text{$2j-1$ is prime}\\[.3cm]
	0 &\text{if}&\begin{subarray}{l} \text{$2j-1$ is a square} \\ \text{of a natural number}\\ \end{subarray}\\[.3cm]
	+\frac{1}{(2j-1)^2} &\text{if}&\begin{subarray}{l}\text{$j=1$ or $2j-1$ is not}\\ \text{a square and not prime}\\\end{subarray}\\
\end{array}\right. .
\end{equation}
For simplicity we shall assume that the cut-off is fixed on $m_c = 13$, so that the only elements of $\mathbf U^{-1}$ contributing to Eq.~\eqref{eq:SASS} come in with the negative sign. Substituting for $a_j$ the exact infinite sums, we obtain
\begin{align}
\nonumber
S^{(\mathrm{AS})}(\omega_p) &= \sum_{j=1}^\infty \frac{S[(2j-1)\omega_p]}{(2j-1)^2}-\sum_{j\in\mathcal J}\frac{S[(2j-1)\omega_p]}{(2j-1)^2}\\
\label{eq:S_AS}
&-\sum_{j\in\mathcal J}\sum_{j'=2}^\infty\frac{S[(2j'-1)(2j-1)\omega_p]}{(2j'-1)^2(2j-1)^2}\,,
\end{align}
where $\mathcal J = \{ 2 ,3 ,4, 6,7\}$ is the set of indices greater than $1$ and no greater than $j_c=(m_c+1)/2$, for which the matrix elements $U_{1,j}^{-1}$ are non-zero.  
The relative error between $S^{(\mathrm{AS})}(\omega_p)$ and the actual value of the spectrum $S(\omega_p)$ is then bounded from above by
\begin{align}
\nonumber
&\frac{|S^{(\mathrm{AS})}(\omega_p)-S(\omega_p)|}{S(\omega_p)} \leqslant \frac{1}{S(\omega_p)}\sum_{\substack{m>m_c\\m\in\mathrm{odd}}}\frac{S(m\omega_p)}{m^2}\\[.3cm]
\nonumber
&\phantom{S(\omega_p)}\leqslant \frac{S[(m_c+2)\omega_p]}{S(\omega_p)}\sum_{\substack{m>m_c\\m\in\mathrm{odd}}}\frac{1}{m^2}\\[.3cm]
\label{eq:AS_error_bound}
&\phantom{S(\omega_p)}
	\leqslant\frac{1}{4}\ln\left(\frac{m_c+2}{2}\right)\frac{S[(m_c+2)\omega_p]}{S(\omega_p)}\,.
\end{align}
The bound shows clearly that the error introduced by the cut-off assumption is mostly insignificant for spectra with finite range, as they decrease much faster than $\ln$ (e.g. for Gaussian spectrum with $\tau_c\,\omega_s=3$ the relative error for $m_c=1$ and $\omega_p = \omega_s$ is of order $10^{-7}$). For long tailed spectra the error might be more significant, but still remains relatively small, as it is illustrated with an example of Lorentzian spectrum in Fig.~\ref{fig:AS_error}.

\begin{figure}[tb]
\centering
\includegraphics[width=\columnwidth]{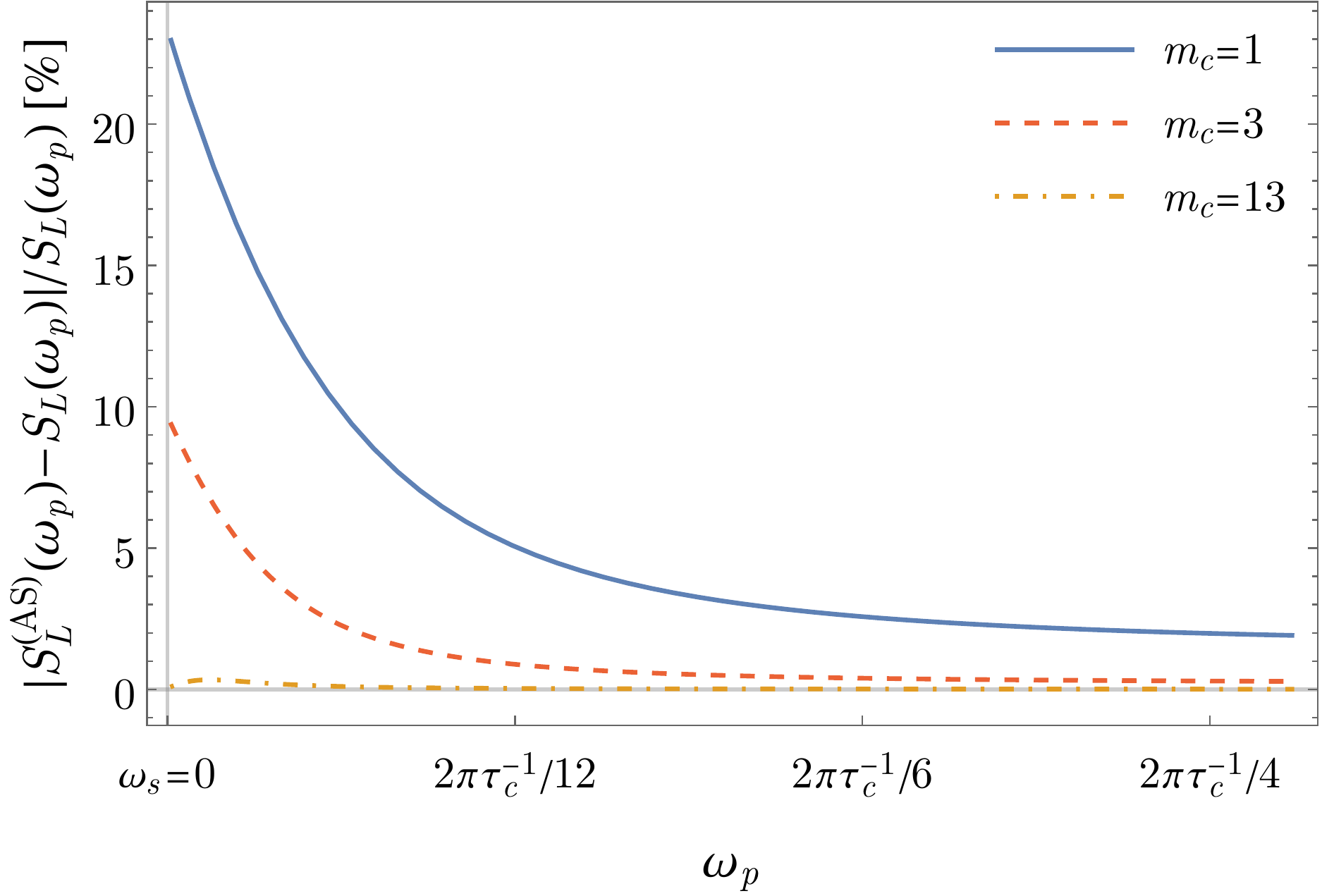}
\caption{The relative error of \as{} method for the reconstruction of a single line Lorentzian spectrum ($\omega_s=0$) with three choices of cut-off $m_c$.}\label{fig:AS_error_0}
\end{figure}
For spectra centered around zero frequency (relevant for Ornstein-Uhlenbeck noise and weak coupling to random telegraph noise, as mentioned before) the bound \eqref{eq:AS_error_bound} ceases to be useful, as it significantly overestimates the actual error. The spectrum of such a noise is a single line Lorentzian, and hence, we can employ the exact formulas for the infinite sums in \eqref{eq:S_AS}, as given by Eqs.~\eqref{eq:CP_single_line_chispec} and \eqref{eq:CP_single_line_dchi}. The result is presented in Fig.~\ref{fig:AS_error_0}, and it shows that even with cut-off set to $m_c=3$, the error is smaller than $10$\%, and drops below $1$\% for larger frequencies.

\section{Conclusion}
We investigated the exact results for the attenuation function of qubit exposed to Gaussian noise for two cases: the long-tailed Lorentzian-shaped and finite-ranged Gaussian-shaped spectra of the noise. Our goal was to better understand the conditions for achieving the spectroscopic regime of pulse sequence parameters, in which the qubit's dephasing rate at time $T$ is related in a simple way to values of the noise spectrum at a discrete set of frequencies determined by the sequence \cite{Alvarez_PRL11}. More formally, the spectroscopic regime is achieved, when the attenuation function $\chi(T)$ from Eq.~\eqref{eq:chi} is well approximated by the spectroscopic formula $\chispec(T)$ from Eq.~\eqref{eq:spec_formula_def}. 

We argued that the correct approach to this problem is to fix the characteristic frequency $\omega_p$ and the Fourier coefficients $\Fc{m\omega_p}$ of the applied pulse sequences. In these settings, the duration of evolution $T$ becomes an adjustable parameter, which can be manipulated by changing the number of repetitions of the chosen pulse sequence. When $T$ is much longer than the noise correlation time, the spectroscopic formula approximates very well the exact attenuation function, but this often comes at a price of making the coherence too small to be reliably measurable. It is thus important to understand the dependence of the correction to the spectroscopic formula, $\dchi$ [see Eq.~\eqref{eq:dchi_def}] on $T$, $\omega_p$ and $\Fc{m\omega_p}$.

We have shown that for the long-tailed Lorentzian spectrum the correction and the spectroscopic formula depend parametrically on $\omega_p$ and $\Fc{m\omega_p}$ in a similar manner, and consequently the spectroscopic regime can be achieved unconditionally by increasing the duration, so that the linear in $T$ scaling of the spectroscopic formula overpowers the $T$-independent correction. In contrast, the finite-ranged Gaussian spectrum decays much faster with increasing $\omega_p$ than the long-tailed correction. Therefore, for characteristic frequencies far from the central frequency of the spectrum, the $T$-overpowering mechanism ceases to be efficient, as it must compensate for rapid decay of the spectral line edges. While these observations have been made by inspecting the exact calculations of the attenuation function for the case of Lorentzian- and Gaussian-shaped spectral densities, the qualitative difference in significance of corrections to the spectroscopic formula between the noise spectra that possess power-law tails, and the ones that are of finite-range, is a general result. The approximate calculations of attenuation function for arbitrary shape of spectral density presented in appendix \ref{app:correction_est}, support this assertion.

The unreliable spectroscopic approximation for finite ranged spectra may lead to misinterpretation of the reconstructed line shape, where the long tail of the correction is incorrectly ascribed to the spectrum of the noise. In other words, an uncritical use of the standard noise spectroscopy method can lead to reconstruction of spectrum with artificial power-law tails, when the real spectrum is in fact finite-ranged. 
We have proposed a scheme that circumvents this problem by exploiting the linear $T$-dependence of the attenuation function for long enough durations. The spectroscopic formula is recovered as a slope of the linear fit to data acquired in a series of measurements with fixed $\omega_p$ and $\Fc{m\omega_p}$ and progressively longer $T$.

Finally, we have discussed the accuracy of \'{A}lvarez-Suter spectrum  reconstruction method \cite{Alvarez_PRL11} applied to data obtained with the modified scheme, and we have derived a simple bound on the error introduced by the assumed frequency cut-off.

\section*{Acknowledgements}
This work is supported by funds of Polish National Science Center (NCN), grant no.~DEC-2015/19/B/ST3/03152.

\onecolumngrid
\appendix
\section{Attenuation function for Lorentzian spectrum}\label{app:lorentz}

Substituting $S(\omega)=(S_\mathrm{L}(\omega-\omega_s)+S_\mathrm{L}(\omega+\omega_s))/2$ and the form \eqref{eq:freq_dom_filter} of the frequency-domain filter function into equation \eqref{eq:attfc_freq}, we can divide all the obtained terms into two groups. First are the ``diagonal'' terms, proportional to $|\Fc{m\omega_p}|^2$, which we rewrite in the following way:
\begin{equation}\label{app:eq:lorentz_diag}
\frac{T^2}{2}\int_{-\infty}^\infty \frac{\id \omega}{2\pi}S(\omega)\,\sinc^2\left(\frac{T(\omega-m\omega_p)}{2}\right)
= \lim_{\eta\to 0^+}\int_{-\infty}^\infty \frac{\id \omega}{2\pi}\frac{S(\omega)(1-e^{i T\omega})}{2(\omega-m\omega_p-i \eta)^2}
	+ \lim_{\eta\to 0^+}\int_{-\infty}^\infty \frac{\id \omega}{2\pi}\frac{S(\omega)(1-e^{-i T\omega})}{2(\omega-m\omega_p-i \eta)^2}\,.
\end{equation}
Here we utilized the relation $m T\omega_p =  2\pi k$ ($k\in\mathrm{Integers}$) and we introduced the {\it convergence factor} $\eta$ which sightly pushes the poles of the integrand into the upper part of the complex plane (the sign of $\eta$ can be chosen arbitrarily). The first integral over real axis can be equated to the line integral over closed semicircle (understood here as an arc \textit{and} the line segment connecting its endpoints) arcing in the upper half of the complex plane $C_+$. Since the Lorentzian spectrum decays as $1/|z|^2$ for large $z$, and $T$ is positive, as the radius of the semicircle is sent to infinity, only the part of contour integral that coincides with the real line remains. Then, the contour integral can be calculated using Cauchy's residue theorem. Denoting by $\mathrm{Res}_{z_0}\left[ S(z)\right]$ the residue of $S(z)$ in point $z_0$, by $z_j^{+} = i\tau_c^{-1}+\omega_s, i\tau_c^{-1}-\omega_s$ the poles of Lorentzian spectrum in the upper half-plane, and by $m\omega_p$ the second order pole of filter function, we proceed with the calculations:
\begin{align}
\nonumber
&\int_{-\infty}^\infty \frac{\id \omega}{2\pi}\frac{S(\omega)(1-e^{i T\omega})}{2(\omega-m\omega_p-i \eta^+)^2}
	= \oint_{C_+}\frac{\id z}{2\pi}\frac{S(z)(1-e^{i T z})}{2(z-m\omega_p-i\eta^+)^2}\\
\nonumber
&= i \sum_{z^+_j}\frac{1-e^{i T z^+_j}}{2(z^+_j-m\omega_p)^2}\mathrm{Res}_{z_j^+}\left[ S(z)\right]
	+ \frac{i}{2}\lim_{z\to m\omega_p}\left[ \frac{\id}{\id z}S(z)(1-e^{i T z})\right]\\
\label{app:eq:Cplus_res}
&=\frac{-i\tau_c^2}{2}\left(\frac{1-e^{-\frac{T}{\tau_c}}e^{i T \omega_s}}{2(1+i\tau_c(m\omega_p-\omega_s))^2}
		+\frac{1-e^{-\frac{T}{\tau_c}}e^{-i T \omega_s}}{2(1+i\tau_c(m\omega_p+\omega_s))^2}\right) \frac{v^2}{i}+\frac{T}{2}S(m\omega_p)\,.
\end{align}
Thus we can see here, that the constituent of the spectroscopic formula, $T/2 S(m\omega_p)$, was obtained as a residue of the filter function. The second term of Eq.~\eqref{app:eq:lorentz_diag} can be calculated in an analogous manner, except the semicircle has to be closed in the lower half-plane (so that the $\exp(-i T z)$ decays when the radius goes to infinity):
\begin{subequations}\begin{align}
\label{subEq:diag_lower:1}
\int_{-\infty}^\infty \frac{\id \omega}{2\pi}\frac{S(\omega)(1-e^{-i T\omega})}{2(\omega-m\omega_p-i \eta^+)^2}
	= \oint_{C_-}\frac{\id z}{2\pi}\frac{S(z)(1-e^{-i T z})}{2(z-m\omega_p-i\eta^+)^2}
	= i \sum_{z^-_j}\frac{1-e^{-i T z^-_j}}{2(z^-_j-m\omega_p)^2}\left(-\mathrm{Res}_{z_j^-}\left[ S(z)\right]\right)&\\
\label{subEq:diag_lower:3}
=\frac{-i\tau^2}{2}\left(\frac{1-e^{-\frac{T}{\tau_c}}e^{-i T \omega_s}}{2(1-i\tau_c(m\omega_p-\omega_s))^2}
		+\frac{1-e^{-\frac{T}{\tau_c}}e^{i T \omega_s}}{2(1-i\tau_c(m\omega_p+\omega_s))^2}\right) \frac{v^2}{i}&\,.
\end{align}\end{subequations}
Here, the poles of spectrum in the lower half-plane are $z_j^{-} = -i \tau_c^{-1}+\omega_s,-i\tau_c^{-1}-\omega_s$, and due to convergence factor $\eta$, the filter function does not contribute any residues.

Having dealt with the diagonal terms, we turn to the second group, the ``off-diagonal'' terms proportional to combinations $\Fc{m_1\omega_p}\Fc{m_2\omega_p}$ with $m_1\neq m_2$:
\begin{align}
&\frac{T^2}{2}e^{i \frac{T(m_1-m_2)\omega_p}{2}}
	\int_{-\infty}^\infty \frac{\id \omega}{2\pi}S(\omega)\,\sinc\left(\frac{T(\omega-m_1\omega_p)}{2}\right)\sinc\left(\frac{T(\omega-m_2\omega_p)}{2}\right)\nonumber\\
&=\int_{-\infty}^\infty \frac{\id \omega}{2\pi}\frac{S(\omega)(1-e^{i T\omega})}{(\omega-m_1\omega_p-i\eta^+)(\omega-m_2\omega_p-i\eta^+)}
	+\int_{-\infty}^\infty \frac{\id \omega}{2\pi}\frac{S(\omega)(1-e^{-i T\omega})}{(\omega-m_1\omega_p-i\eta^+)(\omega-m_2\omega_p-i\eta^+)}\,.
\end{align}
Again, we turn to contour integrals and residue theorem which gives us the following result:
\begin{align}
\nonumber
&\int_{-\infty}^\infty \frac{\id \omega}{2\pi}\frac{S(\omega)(1-e^{i T\omega})}{(\omega-m_1\omega_p-i\eta^+)(\omega-m_2\omega_p-i\eta^+)}
	=\oint_{C_+} \frac{\id z}{2\pi}\frac{S(z)(1-e^{i T z})}{(z-m_1\omega_p-i\eta^+)(z-m_2\omega_p-i\eta^+)}\\
&=i\sum_{z_j^+}\frac{1-e^{i T z_j^+}}{(z_j^+ - m_1\omega_p)(z_j^+-m_2\omega_p)}\mathrm{Res}_{z_j^+}\left[S(z)\right]
	+i \frac{S(m_1\omega_p)(1-e^{i m_1 T \omega_p})}{(m_1-m_2)\omega_p} + i \frac{S(m_2\omega_p)(1-e^{i m_2 T \omega_p})}{(m_2-m_1)\omega_p}\nonumber\\
&=\frac{-i\tau_c^2}{2}\left( \frac{1-e^{-\frac{T}{\tau_c}}e^{i T \omega_s}}{(1+i\tau_c(m_1\omega_p-\omega_s))(1+i\tau_c(m_2\omega_p-\omega_s))} 
	+ \frac{1-e^{-\frac{T}{\tau_c}}e^{-i T \omega_s}}{(1+i\tau_c(m_1\omega_p+\omega_s))(1+i\tau_c(m_2\omega_p+\omega_s))}\right)\frac{v^2}{i}\nonumber\\
&=\left(\int_{-\infty}^\infty \frac{\id \omega}{2\pi}\frac{S(\omega)(1-e^{-i T\omega})}{(\omega-m_1\omega_p-i\eta^+)(\omega-m_2\omega_p-i\eta^+)}\right)^*\,.
\end{align}
Note the contribution form poles of filter function vanish identically because of $m_i T \omega_p = 2\pi k$. Substituting this result back into the off-diagonal sum over $m_1$ and $m_2$, it can be shown (utilizing symmetries of the Fourier coefficients, $c_{-m\omega_p} = c_{m\omega_p}^*$ and by interchanging the names of indices) that together with the diagonal sum, the whole expression can be transformed into form presented in Eqs.~\eqref{eq:Lorentz_full}.

\section{Estimate of the correction to the spectroscopic formula for arbitrary spectral line shape}\label{app:correction_est}

An arbitrary double line spectral density is of the form
\begin{equation}
S(\omega) = \frac{S_0(\omega-\omega_s)+S_0(\omega+\omega_s)}{2} = \int_{-\infty}^\infty \id t\,e^{-i\omega t}C_0(|t|)\cos \omega_s t\,,
\end{equation}
where $\pm\omega_s$ are the line positions (mirrored due to required symmetry of the spectrum), $S_0(\omega)$ is the line shape, and $C(|t|) = C_0(|t|)\cos\omega_s t$ is the corresponding correlation function. Throughout the following calculations we will also assume that the CP pulse sequence was in use, so that $c_{m\omega_p} =e^{im\frac{\pi}{2}} 2/(i\pi m)$.

First we consider the case when the spectrum has a finite range $\tau_c^{-1}$. We will now estimate the value of the correction using the following approximations: (i) we substitute form \eqref{eq:freq_dom_filter} of frequency-domain filter into Eq.~\eqref{eq:attfc_freq} for attenuation function, and (ii) since the line shape has finite range, we can split the sums into far and near peaks terms (see, Sec. \ref{sec:gauss}), (iii) in case of near terms we approximate sinc functions with Dirac deltas, as it was discussed in Sec.~\ref{sec:spec_formula}, Eq.~\eqref{eq:comb_approx}, (iv) in case of far peaks, bearing in mind relation $m_i T \omega_p = 2\pi k$ ($k\in\mathrm{Integers}$), we treat the tails of sinc functions as slowly varying in comparison to spectral density. When we apply those prescriptions we obtain the following
\begin{align}
\nonumber
\chi(T) =&\  \frac{1}{2}\sum_{m_1,m_2}\Fc{m_1\omega_p}\Fc{m_2\omega_p}^*e^{i \frac{T(m_1-m_2)\omega_p}{2}}
	\int_{-\infty}^\infty \frac{\id \omega}{2\pi}S(\omega)\, T\, \sinc\left(\frac{(\omega-m_1\omega_p)T}{2}\right)
	T\, \sinc\left(\frac{(\omega+m_2\omega_p)T}{2}\right) \\
\nonumber
\approx&\  \frac{1}{2}\sum_{m\in\mathrm{near}}|\Fc{m\omega_p}|^2 \int_{-\infty}^\infty \frac{\id \omega}{2\pi}\,S(\omega)\, 2\pi T\,\delta(\omega-m\omega_p)\\
\nonumber
 &+\frac{1}{2}\sum_{\substack{m_1,m_2\in\mathrm{near} \\ m_1\neq -m_2}}\Fc{m_1\omega_p}\Fc{m_2\omega_p}
 	\int_{-\infty}^\infty \frac{\id \omega}{2\pi}S(\omega)\,2\pi T\, \delta(\omega-m_1\omega_p)\,2\pi T\, \delta(\omega+m_2\omega_p)\\
\nonumber
&+\frac{1}{2}\sum_{m_1,m_2\in\mathrm{far}}\frac{\Fc{m_1\omega_p}\Fc{m_2\omega_p}}{(\omega_s-m_1\omega_p)(\omega_s+m_2\omega_p)}
	\frac{1}{2}\int_{-\infty}^\infty \frac{\id \omega}{2\pi}S_0(\omega-\omega_s)(2-e^{i\omega T}-e^{-i\omega T})\\
\nonumber
&+( \mathrm{far}\ :\ \omega_s\to-\omega_s )\\
\nonumber
=&\  \frac{T}{2}\sum_{m\in\mathrm{near}}|\Fc{m\omega_p}|^2 S(m\omega_p) \\
\nonumber
 &+\frac{1}{2}\sum_{m_1,m_2\in\mathrm{far}}\frac{\Fc{m_1\omega_p}\Fc{m_2\omega_p}}{(\omega_s-m_1\omega_p)(\omega_s+m_2\omega_p)}
 	\left[ \int_{-\infty}^\infty \frac{\id \omega}{2\pi} S_\mathrm{0}(\omega) - e^{i \omega_s T}\int_{-\infty}^\infty \frac{\id \omega}{2\pi}S_\mathrm{0}(\omega)e^{i\omega T} \right]\\
\nonumber
&+( \mathrm{far}\ :\ \omega_s\to-\omega_s )\\
\nonumber
\approx&\  T\sum_{m>0}|\Fc{m\omega_p}|^2 S(m\omega_p) \\
\nonumber
&+\frac{1}{2}\sum_{m_1,m_2\in\mathrm{far}}\frac{\Fc{m_1\omega_p}\Fc{m_2\omega_p}}{(\omega_s-m_1\omega_p)(\omega_s+m_2\omega_p)}
	\left[C(0) - e^{i\omega_s T}C_\mathrm{0}(T)\right]\\
\nonumber
&+\frac{1}{2}\sum_{m_1,m_2\in\mathrm{far}}\frac{\Fc{m_1\omega_p}\Fc{m_2\omega_p}}{(-\omega_s-m_1\omega_p)(-\omega_s+m_2\omega_p)}
	\left[C(0) - e^{-i\omega_s T}C_\mathrm{0}(T)\right]\\
\nonumber
=&\  T\sum_{m>0}|\Fc{m\omega_p}|^2 S(m\omega_p) \\
\nonumber
&	+ C(0)\left[\sum_{m\in\mathrm{far}}\frac{\Fc{m\omega_p}}{\omega_s+m\omega_p}\sum_{m'\in\mathrm{far}}\frac{\Fc{m'\omega_p}}{\omega_s-m'\omega_p}\right]
	- C(T)\left[\sum_{m\in\mathrm{far}}\frac{\Fc{m\omega_p}}{\omega_s+m\omega_p}\sum_{m'\in\mathrm{far}}\frac{\Fc{m'\omega_p}}{\omega_s-m'\omega_p}\right]\\
\label{eq:app:finite_range}
\equiv&\  \chispec(T) + \dchi_{0,\mathrm{far}}+\dchi_T(T)\,,
\end{align}
Therefore, our rough estimate shows that the $T$-dependent part of the correction scales as $\sim C(|T|)$, which decays rapidly for $T\gg \tau_c$.

\begin{figure}[h]
\centering
\includegraphics[width=.5\columnwidth]{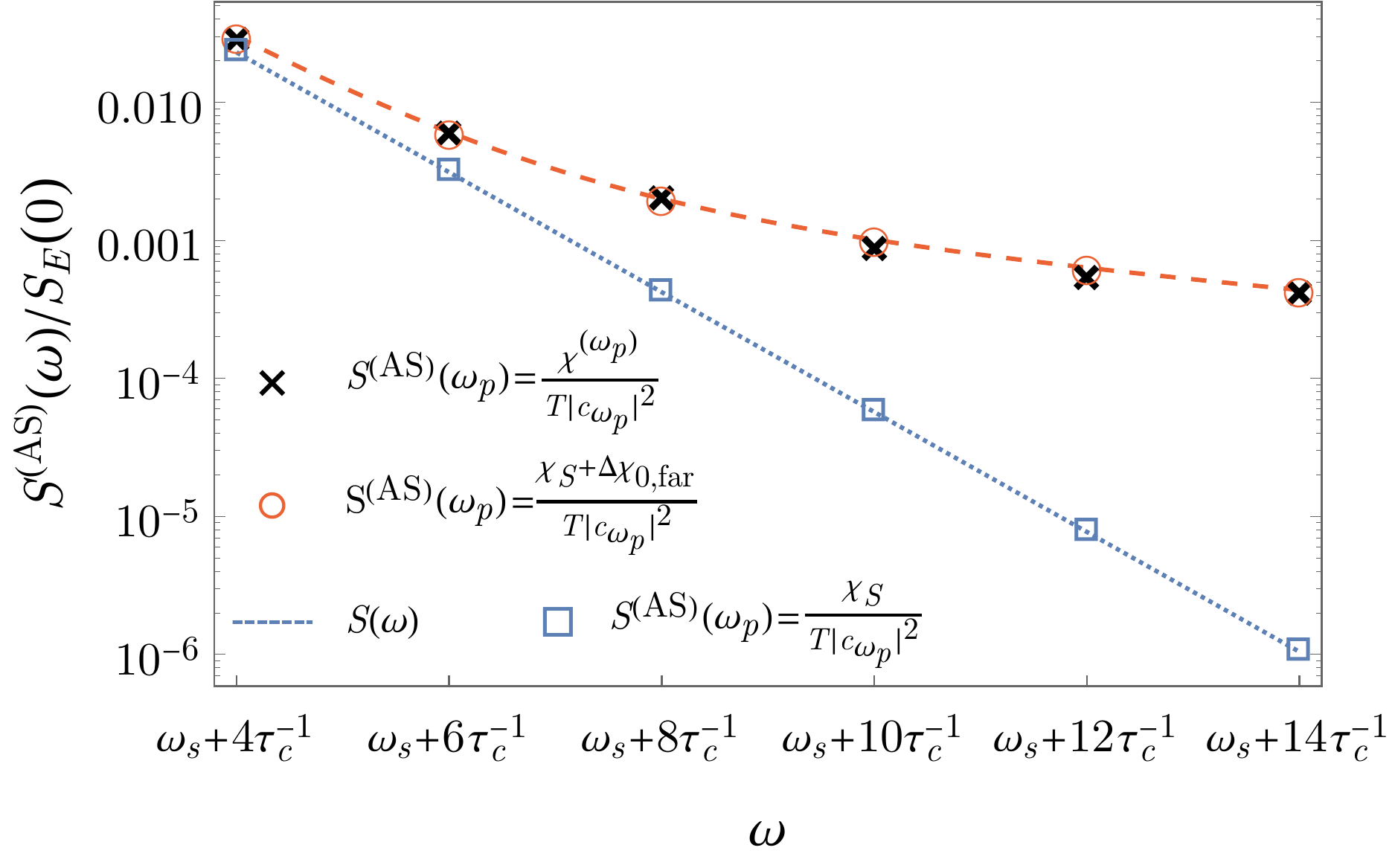}
\caption{An example of exponential spectrum reconstruction with the unmodified \as{} method (the cut-off set to $m_c=1$). The recovered value of spectral density is given by $S^{(\mathrm{AS})}(\omega_p) = \chi^{(\omega_p)}(T)/T|\Fc{\omega_p}|^2$ (crosses). The measured attenuation functions $\chi^{(\omega_p)}(T)$ were simulated via numerical integration of Eq.~\eqref{eq:attfc} using correlation function $C(|t|)=\tau_c\pi^{-1}(\tau_c^2+t^2)^{-1}\cos\omega_s t$ ($\omega_s=2\pi\tau_c^{-1}$) and the time-domain filter functions of an $n=16$ pulse CP sequences with  characteristic frequencies ranging from $\omega_p=\omega_s+4\tau_c^{-1}$ to $\omega_p=\omega_s+14\tau_c^{-1}$. The number of pulses in each sequence was always the same, hence there was no attempt to ramp up the duration in order to compensate for rapid decay of the spectral density. The figure shows on the log scale the spectra $S^{(\mathrm{AS})}$ reconstructed with the exact $\chi^{(\omega_p)}$ (crosses) and the approximate formula \eqref{eq:app:finite_range} (circles), and compares them with the real course of the spectral density $S(\omega)$ (dotted line).}
\label{fig:chi_vs_spec_formula_EXP}
\end{figure}

We verified the veracity of the above approximation by performing numerical experiment where we attempt to use the unmodified \as{} method to reconstruct the tails of exponential spectrum $S(\omega)=[ S_\mathrm{E}(\omega-\omega_s)+S_\mathrm{E}(\omega+\omega_s) ]/2$, $S_\mathrm{E}(\omega)=e^{-\tau_c|\omega|}$ (compare Sec.~\ref{sec:gauss}). The result is presented in Fig.~\ref{fig:chi_vs_spec_formula_EXP}, and it confirms that (i) the estimate for attenuation function is accurate, (ii) similarly to the Gaussian spectrum, the reconstruction of the exponential spectral density faces difficulties with erroneous attribution of long tail behavior.

Second, we consider the case when the spectrum has a long tail, i.e. the line shape $S_0$ decay as $(1/|z|)^\beta$ for $|z|\to\infty$ ($\beta \geqslant 2$ and $\beta\in\mathrm{Integers}$). We can calculate the attenuation function using the same technique of contour integrals as for the Lorentzian spectrum. However, before we proceed with the calculations, it is very helpful to determine the properties of poles and residues of $S(z)$ by examining features of the corresponding correlation function.

For positive $t$, the correlation function can be calculated using the residue theorem by closing the contour of integration in the upper half-plane. Denoting the poles of $S(z)$ enclosed by the contour by $z_j^+ = i\tau_j^{-1} + \omega_j$ and the order of the pole by $r_j$, we obtain the following expression:
\begin{align}
\nonumber
C(t>0)=&\ \int_{-\infty}^\infty \frac{\id\omega}{2\pi}S(\omega)e^{i \omega t} = i \sum_j \mathrm{Res}_{z_j^+=i\tau_j^{-1}+\omega_j}\left[S(z)e^{i z t}\right]\\
\nonumber
=&\ i\sum_j \lim_{z\to i\tau_j^{-1}+\omega_j}\left[ \frac{1}{(r_j-1)!}\frac{\id^{r_j-1}}{\id z^{r_j-1}}e^{i z t}(z-i\tau_j^{-1}-\omega_j)^{r_j}S(z)\right]\\
\nonumber
=&\ \sum_{j}\sum_{k=0}^{r_j-1}t^ke^{-\frac{t}{\tau_j}}e^{i\omega_j t}
	\lim_{z\to i\tau_j^{-1}+\omega_j}\left[i^{k+1}\frac{\id^{r_j-1-k}}{\id z^{r_j-1-k}}\frac{(z-i\tau_j^{-1}-\omega_j)^{r_j}S(z)}{k!(r_j-1-k)!}\right]\\
\equiv&\ \sum_{j}\sum_{k=0}^{r_j-1}t^ke^{-\frac{t}{\tau_j}}e^{i\omega_j t}R_{i\tau_j^{-1}+\omega_j}^{(k)}\,.
\end{align}
The correlation function is Real for every value of $t$, hence
\begin{align}
\nonumber
C(t>0) =&\ \sum_{j}\sum_{k=0}^{r_j-1} t^k e^{-\frac{t}{\tau_j}}\left[ \cos(\omega_j t)\mathrm{Re}\{ R^{(k)}_{i\tau_j^{-1}+\omega_j} \}
	-\sin(\omega_j t)\mathrm{Im}\{ R^{(k)}_{i\tau_j^{-1}+\omega_j} \}\right]\\
\nonumber
&\ +i\sum_{j}\sum_{k=0}^{r_j-1} t^k e^{-\frac{t}{\tau_j}}\left[ \cos(\omega_j t)\mathrm{Im}\{ R^{(k)}_{i\tau_j^{-1}+\omega_j} \}
	+\sin(\omega_j t)\mathrm{Re}\{ R^{(k)}_{i\tau_j^{-1}+\omega_j} \}\right]= C^*(t>0)\,.
\end{align}
Therefore, each summand of the imaginary part of the correlation function must vanish independently, which implies that: (i) the poles come in pairs of the form $i\tau_j^{-1} \pm \omega_j$, and each member of the pair is of the same order. (ii) $R^{(k)}_{i\tau_j^{-1}-\omega_j}=(R^{(k)}_{i\tau_j^{-1}+\omega_j})^*$. Next, we calculate the correlation function for negative times, which requires us to close the contour in the lower half of the complex plane. We parametrize the poles by $z_j^- = -i\mu_j^{-1}+\nu_j$, and $q_j$ is the order of the pole.
\begin{align}
\nonumber
C(t<0)=&\ \int_{-\infty}^\infty \frac{\id\omega}{2\pi}S(\omega)e^{-i \omega |t|} = -i \sum_j \mathrm{Res}_{z_j^-=-i \mu_j^{-1}+\nu_j}\left[S(z)e^{-i z |t|}\right]\\
\nonumber
=&\ -\sum_{j}\sum_{k=0}^{q_j-1}(-|t|)^k e^{-\frac{|t|}{\mu_j}}e^{-i\nu_j |t|}
	\lim_{z\to -i\mu_j^{-1}+\nu_j}\left[i^{k+1}\frac{\id^{q_j-1-k}}{\id z^{q_j-1-k}}\frac{(z+i\mu_j^{-1}-\nu_j)^{q_j}S(z)}{k!(q_j-1-k)!}\right]\\
=&\ \sum_j\sum_{k=0}^{q_j-1}(-|t|)^k e^{-\frac{|t|}{\mu_j}}e^{-i\nu_j |t|}(-R_{-i\mu_j^{-1}+\nu_j}^{(k)})\,.
\end{align}
Again, the requirement that the correlation function is Real implies the poles in the lower half of the complex plane also come in pairs, $-i\mu_j^{-1}\pm\nu_j$, and $R^{(k)}_{-i\mu_j^{-1}-\nu_j} = (R^{(k)}_{-\mu_j^{-1}+\nu_j})^*$. Finally, the correlation function is symmetric, $C(-t)=C(t)$, which implies additional properties: (iii) the poles in the lower and upper half-plane are related: $q_j=r_j$, $\mu_j = \tau_j$, and $\nu_j = \omega_j$, (iv) coefficients $R$ have one more symmetry: $ R^{(k)}_{-i\tau_j^{-1}+\omega_j}=(-1)^{k+1} (R^{(k)}_{i\tau_j^{-1}+\omega_j})^*$.

With all those properties of the residues and poles in hand, we can proceed with the calculation of the attenuation function:
\begin{align}
\nonumber
&\chi(T) =T \sum_{m>0}|\Fc{m\omega_p}|^2 S(m\omega_p) \\ 
\nonumber
&+\frac{1}{2}\sum_{m_1,m_2}\Fc{m_1\omega_p}\Fc{m_2\omega_p}^*\sum_{j}\lim_{z\to z_j^+}\left[i\frac{\id^{r_j-1}}{\id z^{r_j-1}} \frac{1}{(z-m_1\omega_p)(z-m_2\omega_p)}
		\frac{(z-z_j^+)^{r_j} S(z)}{(r_j-1)!}(1-e^{i T z})\right]\\
\nonumber
&-\frac{1}{2}\sum_{m_1,m_2}\Fc{m_1\omega_p}\Fc{m_2\omega_p}^*\sum_{j}\lim_{z\to z_j^-}\left[i\frac{\id^{r_j-1}}{\id z^{r_j-1}} \frac{1}{(z-m_1\omega_p)(z-m_2\omega_p)}
		\frac{(z-z_j^-)^{r_j} S(z)}{(r_j-1)!}(1-e^{-i T z})\right]\\
\nonumber
&=T \sum_{m>0}|\Fc{m\omega_p}|^2 S(m\omega_p) \\
\nonumber
&+\frac{1}{2}\sum_{m_1,m_2}\Fc{m_1\omega_p}\Fc{m_2\omega_p}^*\sum_j\sum_{k=0}^{r_j-1}(-1)^k i^{k}R^{(k)}_{z_j^+}
	\left(\frac{\id^{k}}{\id  z^{k}}\frac{1}{(z-m_1\omega_p)(z-m_2\omega_p)}\Bigg|_{z=z_j^+}\right)\\
\nonumber
&-\frac{1}{2}\sum_{m_1,m_2}\Fc{m_1\omega_p}\Fc{m_2\omega_p}^*\sum_j\sum_{k=0}^{r_j-1}(-1)^k i^{k}R^{(k)}_{z_j^-}
	\left(\frac{\id^{k}}{\id  z^{k}}\frac{1}{(z-m_1\omega_p)(z-m_2\omega_p)}\Bigg|_{z=z_j^-}\right)\\
\nonumber
&-\frac{1}{2}\sum_{m_1,m_2}\Fc{m_1\omega_p}\Fc{m_2\omega_p}^*\sum_j e^{-\frac{T}{\tau_j}}e^{i \omega_j T}\sum_{k=0}^{r_j-1}T^k 
	\lim_{z\to z_j^+}\left[i\frac{\id^{r_j-1-k}}{\id z^{r_j-1-k}}\frac{1}{(z-m_1\omega_p)(z-m_2\omega_p)}\frac{(z-z_j^+)^{r_j}S(z)}{k!(r_j-1-k)!} \right]\\
\nonumber
&+\frac{1}{2}\sum_{m_1,m_2}\Fc{m_1\omega_p}\Fc{m_2\omega_p}^*\sum_j e^{-\frac{T}{\tau_j}}e^{-i \omega_j T}\sum_{k=0}^{r_j-1}T^k 
	\lim_{z\to z_j^-}\left[i\frac{\id^{r_j-1-k}}{\id z^{r_j-1-k}}\frac{1}{(z-m_1\omega_p)(z-m_2\omega_p)}\frac{(z-z_j^-)^{r_j}S(z)}{k!(r_j-1-k)!} \right]\\
&=\chispec(T) + \dchi_0+\dchi_T(T)\,.
\end{align}
As we can see, the $T$-dependent correction decays in similar manner as the correlation function, i.e. as a combination of polynomials in $T$, times exponential function. Therefore, it is justifiable to say that $\dchi_T$ decays as fast as $C(T)$.

To finish our analysis, we shall examine the behavior of the $T$-independent correction $\dchi_0$ as a function of characteristic frequency $\omega_p$. In particular, we seek to estimate the behavior of the correction in the limit of large $\omega_p$.
\begin{subequations}
\begin{align}
\nonumber
\dchi_0 =&\  
	\frac{1}{2}\sum_{m_1,m_2}\Fc{m_1\omega_p}\Fc{m_2\omega_p}^*\sum_j\sum_{k=0}^{r_j-1}(-1)^k i^k R^{(k)}_{z_j^+}
		\left(\frac{\id^{k}}{\id  z^{k}}\frac{1}{(z-m_1\omega_p)(z-m_2\omega_p)}\Bigg|_{z=z_j^+}\right)\\
\nonumber
&-\frac{1}{2}\sum_{m_1,m_2}\Fc{m_1\omega_p}\Fc{m_2\omega_p}^*\sum_j\sum_{k=0}^{r_j-1}(-1)^k i^k R^{(k)}_{z_j^-}
		\left(\frac{\id^{k}}{\id  z^{k}}\frac{1}{(z-m_1\omega_p)(z-m_2\omega_p)}\Bigg|_{z=z_j^-}\right)\\
\nonumber
\approx&\ 
	\frac{1}{2}\sum_{m_1,m_2}\Fc{m_1\omega_p}\Fc{m_2\omega_p}^*\sum_j \Bigg[ 
		R^{(0)}_{i\tau_j^{-1}+\omega_j}\frac{1}{(i\tau_j^{-1}+\omega_j-m_1\omega_p)}\frac{1}{(i\tau_j^{-1}+\omega_j-m_2\omega_p)}\\
\nonumber
&\ \phantom{\frac{1}{2}\sum_{m_1,m_2}\Fc{m_1\omega_p}\Fc{m_2\omega_p}^*\sum_j }
		+R^{(0)}_{i\tau_j^{-1}-\omega_j}\frac{1}{(i\tau_j^{-1}-\omega_j-m_1\omega_p)}\frac{1}{(i\tau_j^{-1}-\omega_j-m_2\omega_p)}\\
\nonumber
&\ \phantom{\frac{1}{2}\sum_{m_1,m_2}\Fc{m_1\omega_p}\Fc{m_2\omega_p}^*\sum_j }
		-R^{(0)}_{-i\tau_j^{-1}+\omega_j}\frac{1}{(-i\tau_j^{-1}+\omega_j-m_1\omega_p)}\frac{1}{(-i\tau_j^{-1}+\omega_j-m_2\omega_p)}\\
\label{subEq:LT_dchi0_k_is_0}
&\ \phantom{\frac{1}{2}\sum_{m_1,m_2}\Fc{m_1\omega_p}\Fc{m_2\omega_p}^*\sum_j }
		-R^{(0)}_{-i\tau_j^{-1}-\omega_j}\frac{1}{(-i\tau_j^{-1}-\omega_j-m_1\omega_p)}\frac{1}{(-i\tau_j^{-1}-\omega_j-m_2\omega_p)}\Bigg]\\
\nonumber
=&\ -\sum_j \tau_j^2 R^{(0)}_{i\tau_j^{-1}+\omega_j}
	\sum_{m}\frac{\Fc{m\omega_p}}{1+i\tau_j(m\omega_p-\omega_j)}
	\sum_{m'}\frac{\Fc{m'\omega_p}^*}{1+i\tau_j(m'\omega_p-\omega_j)}\\
\label{subEq:LT_dchi0_structure}
&\ -\sum_j \tau_j^2 (R^{(0)}_{i\tau_j^{-1}+\omega_j})^*
	\sum_{m}\frac{\Fc{m\omega_p}}{1+i\tau_j(m\omega_p+\omega_j)}
	\sum_{m'}\frac{\Fc{m'\omega_p}^*}{1+i\tau_j(m'\omega_p+\omega_j)}\\
\label{subEq:LT_dchi0}
\xrightarrow{\omega_p\gg\omega_j}&\ 
	\sim -2\sum_j \frac{\tau_j^2 \,\mathrm{Re}\left\{ R^{(0)}_{i\tau_j^{-1}+\omega_j}\right\}}{(\tau_j \omega_p)^4}
\end{align}
\end{subequations}
Since $\id^{k}/\id z^{k} [(z-m_1\omega_p)(z-m_2\omega_p)]^{-1} \sim 1/\omega_p^{2+k}$, and we are looking for the term that has the longest tail, in Eq.~\eqref{subEq:LT_dchi0_k_is_0} we truncated the sum over $k$ at the first term. Then, in Eq.~\eqref{subEq:LT_dchi0_structure} we showed that the structure of the sums over harmonics of the filter function ($m,m'$) is identical to those found in the correction in the case of Lorentzian spectrum (see, Eq.~\eqref{subEq:Lorentz_full_dchi}). This leads to the estimate of large $\omega_p$ behavior listed in \eqref{subEq:LT_dchi0}.
\twocolumngrid

\end{document}